\documentclass[12pt]{article}%

\usepackage{amssymb}
\usepackage{amsfonts}
\usepackage{amsmath}
\usepackage{natbib}
\usepackage[nohead]{geometry}
\usepackage[singlespacing]{setspace}
\usepackage[bottom]{footmisc}
\usepackage{indentfirst}
\usepackage{endnotes}
\usepackage{graphicx}%
\usepackage{rotating}
\usepackage{verbatim}
\usepackage{setspace}
\usepackage{multirow}
\usepackage{latexsym}

\usepackage{amsthm}
\usepackage{makeidx}
\usepackage{fancyhdr}
\usepackage{type1cm}

\newcommand{\fcs}{\text{FCS}}

\newcommand{\bcs}{\text{BCS}}
\newcommand{\sgn}{\text{sgn}}

\newcommand{\var}{\mathrm{var}}
\newcommand{\beq}{\begin{eqnarray*}}
\newcommand{\eeq}{\end{eqnarray*}}

\newtheorem{thm}{Theorem}[section]

\newtheorem{lem}{Lemma}[section]
\newtheorem{cor}{Corollary}[section]
\newtheorem{assum}{Assumption}[section]

\numberwithin{equation}{section}

\theoremstyle{definition}
\newtheorem{exm}{Example}[section]
\newtheorem{remark}{Remark}[section]

\usepackage{textcomp}
\usepackage{xr}
\makeatletter
\def\@biblabel#1{\hspace*{-\labelsep}}
\makeatother
\begin{document}

\title{Semi-parametric Bayesian Partially Identified Models based on Support Function\thanks{The authors are grateful to Federico Bugni,  Ivan Canay,   Joel Horowitz, Enno Mammen, Francesca Molinari, Andriy Norets, Adam Rosen, Frank Schorfheide, J\"{o}rg Stoye and seminar participants at CREST (LS and LMI), Luxembourg,  Mannheim, Tsinghua, THEMA, University of Illinois at Urbana-Champaign, $4^{th}$ French Econometrics Conference, Bayes in Paris workshop, Oberwolfach workshop, 2013 SBIES meeting, CMES  2013, EMS 2013 in Budapest and ESEM 2013 in Gothenburg for useful comments. Anna Simoni gratefully acknowledges financial support from the University of Mannheim through the DFG-SNF Research Group FOR916, ANR-13-BSH1-0004, labex MMEDII (ANR11-LBX-0023-01) and hospitality from University of Mannheim and CREST.  }}
\author{Yuan Liao\footnote{Department of Mathematics, University of Maryland at College Park, College Park, MD 20742 (USA).   Email: yuanliao@umd.edu}\medskip\\{University of Maryland}  \and Anna Simoni\footnote{CNRS and THEMA, Universit\'{e} de Cergy-Pontoise - 33, boulevard du Port, 95011 Cergy-Pontoise (France). Email: simoni.anna@gmail.com}\medskip\\{CNRS and THEMA}  }

\date{November 2013}
\maketitle
\begin{abstract}

We provide a comprehensive semi-parametric study of Bayesian partially identified econometric models. While the existing  literature on Bayesian partial identification has mostly focused on the structural parameter, our   primary focus is on Bayesian credible sets (BCS's) of the unknown identified set and the posterior distribution of its support function. We construct a (two-sided) BCS  based on the support function of the identified set.   We prove the Bernstein-von Mises theorem for the posterior distribution of the support function.   This powerful result in turn infers that, while the BCS  and the frequentist confidence set  for the partially identified parameter are asymptotically different, our constructed BCS for the identified set has an asymptotically correct frequentist coverage probability. Importantly, we illustrate that the constructed  BCS for the identified set does not require a prior on the structural parameter. It can be computed efficiently for subset inference, especially when the target of interest is a sub-vector of the partially identified parameter, where projecting to a low-dimensional subset is often required. Hence, the proposed methods are  useful in many applications.

The Bayesian  partial identification literature has been assuming  a known parametric likelihood function. However,  econometric models usually only identify a set of moment inequalities, and therefore using  an incorrect likelihood function  may result in misleading inferences. In contrast, with a nonparametric prior on the unknown likelihood function, our proposed Bayesian procedure only requires a set of moment conditions, and can efficiently make inference about both the partially identified parameter and its identified set.  This makes it widely applicable in general  moment inequality models.  Finally, the proposed method is  illustrated in a financial asset pricing problem.

\end{abstract}


\newpage
\section{Introduction}

Partially identified models have been receiving extensive attentions in recent years due to their broad applications in econometrics. Partial identification of a structural parameter arises when the data available and the constraints coming from economic theory only allow to place the parameter inside a proper subset of the parameter space. Due to the limitation of the data generating process, the data cannot provide any information within the set where the structural parameter is partially identified (called \textit{identified set}).

This paper aims at developing a semi-parametric Bayesian inference for partially identified models. A Bayesian approach may be appealing for several reasons. First, Bayesian procedures often conduct inference through Bayesian credible sets (BCS's), which are often relatively easy to construct thanks to the use of Markov Chain Monte Carlo (MCMC) methods. This is particularly useful when we are concerned about marginalizing the BCS to a low-dimensional space.  In some situations,   we are interested only in a projection of the identified region for  the subset inference.   We demonstrate that     our proposed approach provides tractable computational tools for projecting a high-dimensional identified region to a low-dimensional space. This   has important implications for practical implementations.

Secondly, our Bayesian procedures also have comprehensive frequentist validations. In particular, our constructed   BCS of the identified set also has a correct asymptotic frequentist coverage probability. We construct credible sets  based on the support function; the latter completely characterizes convex and closed identified sets. We also show the Bernstein-von Mises theorem for the posterior distribution of the support function.  At the best of our knowledge this has not been studied in the literature yet.  This powerful result in turn allows us to establish the (asymptotic) equivalence between BCS's and frequentist confidence sets (FCS's) for the identified set. The literature on partial identification distinguishes between credible/confidence sets for the partially identified parameter and for the identified set. Credible sets for the identified set play an important role not only when the target of interest is the partially identified parameter but even when the identified set is itself the object of interest. While focusing on the study of BCS for the identified set, we also extend  Moon and Schorfheide (2010)'s analysis for the partially identified parameter to a semi-parametric setup,  which is relevant in more general \textit{moment inequality models} where the likelihood function may be unknown. Moreover,  if we admit the existence of a true value of the structural parameter and the identified set,  the corresponding posterior distributions concentrate asymptotically in a neighborhood of the true value (set). This property is known as the \textit{posterior consistency}. It is important because it guarantees that, with a sufficiently large amount of data, we can recover the truth accurately with large probabilities. 

Third, putting a prior on the partially identified parameter can be viewed  as a way of incorporating researchers' beliefs. A Bayesian approach conveniently combines the information from both the observed data and other sources of prior information. The prior information is, for instance, information coming from historical data, information based on experience or on previous survey data. In some applications this information is largely available, \textit{e.g.} in macroeconomics, central banking and finance.   We stress that the prior information will not affect the boundary of the identified set, but will only play a role in determining which areas inside the identified set are a priori ``more likely'' than others.   On the other hand, when specifying a prior distribution on the partially identified parameter is either  difficult or conflicting with the philosophy of partial identification, a  researcher can still use our procedure and either specify a uniform prior or just construct the BCS for the identified set for inference.   The latter is due to an important feature   of our procedure  that the Bayesian analysis of the identified set does not  require to specify a prior on the partially identified parameter.   Therefore,  we accommodate both situations where a researcher does  have prior information as well as situations where she does not.

From the posterior perspective, the Bayesian  partial identification produces a posterior distribution of the partially identified parameter whose support will asymptotically concentrate around the true identified set.  When informative  priors are available,  the shape of the posterior density may  not be flat   inside the identified set,  and  will ground on the prior distribution even asymptotically. Therefore, the asymptotic behavior of the posterior distribution is different from that of the traditional point identified case where (in the latter case) the information from the prior is often washed away by the data asymptotically. Thus, the Bayesian approach to partial identification links conclusions and inferences to various information sources -- data, prior, experience, etc.-- in a transparent way.

Finally, when the identified set depends on a point identified nuisance parameter, say $\phi$, and this is integrated out with respect to its posterior, then the prior information on the partially identified parameter is completely revised by the data. Hence,  the proposed procedure also learns about the partially identified parameter based on the whole posterior distribution of $\phi$, which is potentially  useful  in finite samples. Consequently, there is a strong  motivation for us to conduct a comprehensive Bayesian study for the partially identified econometric models.\\

There are in general two approaches in the literature on Bayesian partial identification. The first approach specifies a parametric likelihood function and assumes it is known  up to a finite-dimensional parameter. This approach has been used frequently in the literature, see e.g., Moon and Schorfheide (2012), Poirier (1998),  Bollinger and  Hasselt (2009), Norets and Tang (2012) among many others. In many applications, however, econometric models usually only identify a set of moment inequalities instead of the full likelihood function. Examples are: interval-censored data, interval instrumental regression, asset pricing (Chernozhukov et al. 2008),  incomplete structural models (Menzel 2011), etc. Assuming a parametric form of the likelihood function is  ad-hoc in these applications. Once the likelihood is mis-specified, the posterior can be misleading. The second approach starts from a set of moment inequalities, and uses a moment-condition-based likelihood such as the limited information likelihood (Kim 2002) and the exponential tilted empirical likelihood (Schennach 2005). Further references may be found in Liao and Jiang (2010),  Chernozhukov and Hong (2003) and Wan (2011). This approach avoids assuming the knowledge of the true likelihood function. However, it only studies the structural parameter, and it is hard to construct posteriors and credible sets for the identified set. Moreover, it does not have a Bayesian probabilistic interpretation.

This paper proposes a pure Bayesian procedure without assuming a parametric form of the true likelihood function. We place a nonparametric prior on the likelihood and obtain the marginal posterior distribution for the partially identified parameter  and  the identified set. A similar Bayesian procedure was recently used in Florens and Simoni (2011). As a result, our procedure is semi-parametric Bayesian and  can make inference about both the partially identified parameter and its identified set easily. It only requires a set of moment conditions and then it can be completely nonparametric on the data generating process.  This is an appealing feature in general moment inequality models.   On the other hand, if the likelihood function is known,   our procedure continues to work and this paper is still well-motivated. In fact, many contributions of this paper, e.g., Bayesian inference of the  support function, construction of BCS for the identified set, subset inferences, etc., are relevant and original also for the case with a known likelihood.



There is a growing literature on Bayesian partially identified models. Besides those mentioned above, the list also includes Gelfand and Sahu (1999), Neath and Samaniego (1997), Gustafson (2012), Epstein and Seo (2011), Stoye (2012), Kitagawa (2012), Kline (2011), etc. There is also an extensive literature that analyzes partially identified models from a frequentist point of view. A partial list includes Andrews and Guggenberger (2009), Andrews and Soares (2010), Andrews and Shi (2013), Beresteanu, Molchanov and Molinari (2011), Bugni (2010), Canay (2010), Chernozhukov, Hong and Tamer (2007),  Chiburis (2009),  Imbens and Manski (2004), Romano and Shaikh (2010),  Rosen (2008),  Stoye (2009),  among others. See Tamer (2010) for a review.

When the identified set is closed and convex, the support function becomes one of the useful tools to characterize its properties.  The literature on this perspective has been growing rapidly, see for example,   Bontemps, Magnac and Maurin (2012),  Beresteanu and Molinari (2008),   Beresteanu et al. (2012),   Kaido and Santos (2013),  Kaido (2012) and Chandrasekhar et al. (2012). This   paper is also closely related to the asymptotic  nonparametric Bayesian literature: Wu and Ghosal (2008), Ghosh and Ramamoorthi (2003), Ghosal and van der Vaart (2001),  Shen and Wasserman (2001),  Ghosal et al. (1999), Amewou-Atisso et al. (2003),   Walker et al. (2007),   van der Vaart and van Zanten (2008), Bickel and Kleijn (2012),  Jiang (2007), Choi and  Ramamoorthi (2008), Castillo (2008),  Freedman (1999),  Rivoirard  and Rousseau (2012),  among others.


The paper is organized as follows. Section \ref{highlight} outlines our main results and contributions. Section \ref{s_general_setup} sets up the model and discusses the prior specification on the underlying likelihood function. Section \ref{s_nonparametric_prior} studies the (marginal) posterior distribution of the structural parameter. Section \ref{ssf} studies the posterior of the support function in moment inequality models. In particular, the Bernstein-von Mises theorem and a linear representation for the support function are obtained. Section \ref{sbcs} constructs the Bayesian credible sets for both the structural parameter and its identified set. In addition, the frequentist coverages of these credible sets are studied.  Section \ref{s_Projection_Subset_Inference} addresses the subset inference when the target of interest is only a component of the full parameter. Section \ref{s_posterior_consistency_set} shows the posterior consistency for the identified set and provides the concentration rate. Section \ref{s_further_illustration_Uniformity} addresses the uniformity. In particular, it discusses the case when point identification is actually achieved.  Section \ref{s_Financial_asset_pricing} applies the support function approach to a financial asset pricing study. 
Finally, Section \ref{s_Conclusion} concludes with further discussions. All the proofs are given in the appendix to this paper and in a supplementary appendix.

\section{Highlights of Our Contributions}\label{highlight}

This section provides a global vision of our main contributions of this paper. Formal setup of the model starts from Section 3.


\subsubsection*{Semi-parametric Bayesian partial identification}

We focus on semi-parametric models where the true likelihood function may be unknown, which is more relevant in moment inequality models. Then there are three types of parameters in the Bayesian setup: $\theta$, which is the partially identified structural parameter;  $\phi$,   a   point-identified parameter that characterizes the identified set, and the  unknown likelihood  $F$.  The identified set can be written as $\Theta(\phi)$. According to the Bayesian philosophy, we treat the identified set as random, and construct its posterior distribution. 

Without assuming any parametric form for the likelihood, we place a nonparametric prior $\pi(F)$ on it. The posteriors of $\phi$ and of the identified set can then be constructed via the posterior of $F$.  Such a semi-parametric posterior requires only a set of moment inequalities, and therefore is robust to the likelihood specification.  Moreover, to make inference about the partially identified $\theta,$ we place a conditional prior $\pi(\theta|\phi)$ supported only on $\Theta(\phi)$.  Note that Bayesian inference for the identified set may be carried out based on the posterior of $\Theta(\phi)$ which does not depend on $\pi(\theta|\phi)$. So the  prior specification for $\theta$ plays a role only in the inference about $\theta$.



For these posteriors, we show that asymptotically $p(\theta|Data)$ will be supported  within an arbitrarily small neighborhood of the true identified set, and the  posterior of $\Theta(\phi)$ also concentrates around the true set in the Hausdorff distance. These are  the notion of \emph{posterior consistency} under partial identification. 

\subsubsection*{Support function}

To make inference on $\Theta(\phi)$ we can take advantage of the fact that when $\Theta(\phi)$ is closed and convex it is completely characterized by its \textit{support function} $S_{\phi}(\cdot)$ defined as:
$$
S_{\phi}(\nu)=\sup_{\theta\in\Theta(\phi)}\theta^T\nu
$$
where $\nu\in\mathbb{S}^{\dim(\theta)}$, the unit sphere. Therefore, inference on $\Theta(\phi)$ may be conveniently carried out through inference on its support function. The posterior distribution of $S_{\phi}(\cdot)$ is also determined by that of $\phi$. We show that in a general moment inequality model, the support function has an asymptotic linear representation in a neighborhood of the true value of $\phi$, which  potentially extends the inference in Bontemps et al. (2012) to nonlinear models.
Our paper also establishes  the Bernstein-von Mises theorem for the support function, that is, the posterior distribution of  $S_{\phi}(\cdot)$  converges weakly to a Gaussian process.  
We also calculate the support function for a number of interesting examples, including interval censored data, missing data, interval instrumental regression and asset pricing models.


\subsubsection*{Two-sided Bayesian credible sets for the identified set}

We construct two types of Bayesian credible sets (BCS's): one for the identified set $\Theta(\phi)$ and the other for the partially identified parameter $\theta$. In particular, the BCS for the  {identified set} is constructed based on the support function, is two-sided, and has an asymptotically correct frequentist coverage probability.  Specifically,  we find sets $\Theta(\hat\phi )^{-q_{\tau}/\sqrt{n}}$ and $\Theta(\hat\phi )^{q_{\tau}/\sqrt{n}}$,  satisfying:  for  level $1-\tau$ where $\tau\in (0,1)$,

Bayesian coverage:
\begin{equation}\label{eq2.1add}
P(\Theta(\hat\phi )^{-q_{\tau}/\sqrt{n}}\subset\Theta(\phi)\subset\Theta(\hat\phi )^{q_{\tau}/\sqrt{n}}|Data)=1-\tau;
\end{equation}

Frequentist coverage:
\begin{equation}\label{eq2.2add}
P_{D_n}(\Theta(\hat\phi )^{-q_{\tau}/\sqrt{n}}\subset\Theta(\phi_0)\subset\Theta(\hat\phi )^{q_{\tau}/\sqrt{n}})\geq1-\tau,
\end{equation}
where   $P_{D_n}$ denotes the sampling probability, and $\Theta(\phi_0)$ is the true identified set.    In (\ref{eq2.1add}) the random set is $\Theta(\phi)$ while in (\ref{eq2.2add}) the random sets are $\Theta(\hat\phi )^{-q_{\tau}/\sqrt{n}}$ and $\Theta(\hat\phi )^{q_{\tau}/\sqrt{n}}$.  One of the important features is that the BCS for the identified set does not require specifying a prior on the partially identified parameter.  The notation $\Theta(\hat\phi )^{-q_{\tau}/\sqrt{n}}$, $\Theta(\hat\phi )^{q_{\tau}/\sqrt{n}}$, $\hat\phi$ and $q_{\tau}$ are to be formally defined in the paper. Therefore, the constructed two-sided BCS can also be used as frequentist confidence set for the identified set.

Furthermore, we find that in the semi-parametric Bayesian model, Moon and Schorfheide (2012)'s conclusion about the BCS for the partially identified parameter $\theta$ still holds:  it is smaller than frequentist confidence sets in large samples.
Hence, while the BCS for the partially identified parameter does not have a correct frequentist coverage, the asymptotic equivalence between BCS and FCS for the identified set holds. Intuitively, this is because the prior information still plays an important role in the posterior of the partially identified parameter even asymptotically; on the other hand, as the identified set is ``point identified", whose BCS is independent of the prior on $\theta$, then its prior information is ``washed away'' asymptotically.  Thus, the proposed inference for the identified set and the support function is asymptotically robust to their prior specification.

\subsubsection*{Projection and subset inference}

We show that with our approach it is easy to project 
(marginalize) onto low-dimensional subspaces for subset inferences. This computation is fast. 
Suppose the dimension of $\theta$ is relatively large, but we are interested in only a few components of $\theta$,   and aim to make inference about these components and their marginal identified set. In our approach, constructing the identified set and BCS for the marginal components   simply requires the marginalization of a joint distribution and can be carried out efficiently thanks to the use of MCMC methods. It is also computationally convenient to calculate the BCS for the marginal identified set. Hence, the proposed   procedure has large potentiality in many empirical applications.

\subsubsection*{Uniformity}

The proposed  Bayesian inference for the identified set is valid uniformly over a class of data generating process. In particular, using specific examples, we illustrate that as the identified set shrinks to a singleton, so that point identification is (nearly) achieved, our Bayesian inference for the identified set carries over.

\subsubsection*{Applications}

We develop a detailed application of Bayesian partial identification to financial asset pricing, which is an example where the identified set is of direct interest. Estimation and inference for the support function as well as for the identified set are conducted. Moreover, throughout the paper, we study in detail other typical examples including the interval censoring, interval regression and missing data problems. 


\section{General Setup of Bayesian Partially Identified Model}\label{s_general_setup}

\subsection{The Model}
Econometric models often involve a structural parameter $\theta\in\Theta$ that is only partially identified by the data generating process (DGP) on a non-singleton set, which we call \textit{identified set}. The model also contains two parameters that are point identified by the DGP:  a finite-dimensional parameter $\phi\in\Phi\subset\mathbb{R}^{d_{\phi}}$ and the distribution function $F$ of the observed data, which is infinite-dimensional. Here, $\Phi$ denotes the parameter space for $\phi$ and $d_{\phi}$ its dimension. The point identified parameter $\phi$ often arises naturally as it characterizes the data distribution. In most of partially identified models, the identified set is also characterized by $\phi$, hence we denote it by $\Theta(\phi)$ to indicate that once $\phi$ is determined, so is the identified set. Let $d=\dim(\theta)$ and $\Theta\subset\mathbb{R}^{d}$ denote the parameter space for $\theta$; we assume $\Theta(\phi)\subseteq\Theta$.

In a parametric Bayesian partially identified model as in  Poirier (1998), Gustafson (2012) and Moon and Schorfheide (2012), $F$ is linked with a known likelihood function to $\phi$. However, as in the usual point identified models, in some applications  assuming a known likelihood function may suffer from a model specification problem, and may lead to misleading conclusions. Instead, econometric applications often  involve only a set of moment conditions as in (\ref{eq2.1}) below. This gives rise  to the \textit{moment inequality models}.  A parametric form of the likelihood function and of $F$ can be unavailable in these models. A robust approach is to proceed without assuming a parametric form for the likelihood function, but to put a prior on $(\theta,\phi,F)$ instead.   This yields the semi-parametric Bayesian setup.

We  specify a nonparametric prior  on data's cumulated distribution function (CDF) $F$, which can deduce a prior for $\phi $ through a transformation $\phi=\phi(F)$, as $\phi$ often is a functional of $F$.   Moreover, the prior  on the identified set $\Theta(\phi)$   is determined through that of $\phi$.
 Due to the identification feature, for any given $\phi\in\Phi$, we specify a conditional prior $\pi(\theta|\phi)$ such that
$$
\pi(\theta\in\Theta(\phi)|\phi)=1.
$$
By construction, this prior for $\theta$ puts all its mass on $\Theta(\phi)$ for any $\phi\in\Phi$.  So it takes the form:
    \begin{displaymath}
      \pi(\theta|\phi) \propto I_{\theta\in\Theta(\phi)}g(\theta),
    \end{displaymath}
where $g(\cdot)$ is some probability density function and $I_{\theta\in\Theta(\phi)}$ is the indicator function of $\Theta(\phi)$. In Section \ref{app_prior_theta} we discuss the philosophy of specifying the prior on $\theta$.

Our analysis focuses on the situation where $\Theta(\phi)$ is a closed and convex set for each $\phi$. Therefore, $\Theta(\phi)$ can be uniquely characterized by its \emph{support function}. For any fixed $\phi$, the support function for $\Theta(\phi)$ is a function $S_{\phi}(\cdot): \mathbb{S}^d\rightarrow\mathbb{R}$ such that
    \begin{displaymath}
      S_{\phi}(\nu)= \sup_{\theta\in\Theta(\phi)}\theta^T\nu.
    \end{displaymath}
where $\mathbb{S}^d$ denotes the unit sphere in $\mathbb{R}^d$. The support function plays a central role in convex analysis since it determines all the characteristics of a convex set. Hence, it is one of the essential objects for our Bayesian inference. In a similar way as for $\Theta(\phi)$, we put a prior on $S_{\phi}(\cdot)$ via the prior on $\phi$.

Suppose $p(\phi|D_n)$ denotes the posterior of $\phi$, given the data $D_n$ and a prior $\pi(\phi)$. It is readily seen  that (see e.g., Poirier 1998) the joint posterior of $(\theta, \phi)$ is given by
$$
p(\theta, \phi|D_n)\propto\pi(\theta|\phi)p(\phi|D_n).
$$
By integrating out $\phi$, we obtain the marginal posterior for $\theta$.  On the other hand,  the posteriors of $\Theta(\phi)$ and $S_{\phi}(\cdot)$ are also determined through the marginal posterior  $p(\phi|D_n)$.   This also highlights an important feature of this paper: our results  on $\Theta(\phi)$ and the support function  do not require placing a prior on the partially identified parameter $\theta$,  because  as far as $p(\phi|D_n)$ is concerned, the prior for $\theta$ is not needed at all.  Furthermore, as the identified set and support function are ``point identified'', their posteriors are asymptotically robust to the prior specifications on $\phi$.


Let us present a few examples that have received much attention in partially identified econometric models literature. In the rest of the paper, we denote by $X$ the observable random variable for which we have $n$ i.i.d. observations $D_n=\{X_i\}_{i=1}^n$. Let $(\mathcal{X},\mathcal{B}_{x},F)$ denote a probability space in which $X$ takes values and $\mathcal{F}$ denote the parameter space of $F$.

\begin{exm}[Interval censored data]\label{ex2.1}
  Let $(Y,Y_{1},Y_{2})$ be a $3$-dimensional random vector such that $Y\in[Y_{1},Y_{2}]$ with probability one. The random variables $Y_{1}$ and $Y_{2}$ are observed while $Y$ is unobservable (see, e.g., Moon and Schorfheide 2012).  We denote:  $\theta=E(Y)$ and $\phi=(\phi_{1},\phi_{2})' \equiv (E(Y_{1}),E(Y_{2}))'$. Therefore, we have the following
identified set for $\theta$: $\Theta(\phi)=[\phi_1, \phi_2]$. The support function for $\Theta(\phi)$ is easy to derive:
    \begin{displaymath}
   S_{\phi}(1)=\phi_2, \quad S_{\phi}(-1)=-\phi_1.
    \end{displaymath}
    The non-parametric prior specification on the likelihood is to be discussed in Section \ref{sss_nonparametric_prior}.
$\square$


\end{exm}

\begin{exm}[Interval regression model]\label{ex2.2}
 The regression model with interval censoring has been studied by, for example, Haile and Tamer (2003). Let $(Y,Y_{1},Y_{2})$ be a $3$-dimensional random vector such that $Y\in[Y_{1},Y_{2}]$ with probability one. The random variables $Y_{1}$ and $Y_{2}$ are observed while $Y$ is unobservable. Assume that
$$
Y=x^T\theta+\epsilon
$$
where $x$ is a vector of observable regressors. In addition, assume there is a $d$-dimensional vector of nonnegative exogenous variables $Z$ such that $E(Z\epsilon)=0$. Here $Z$ can be either a vector of instrumental variables when $X$ is endogenous, or a nonnegative transformation of $x$ when $x$ is exogenous. It follows that
\begin{equation}\label{eq2.1}E(ZY_1)\leq E(ZY)=E(Zx^T)\theta\leq E(ZY_2).
\end{equation}
We denote $\phi=(\phi_1, \phi_2, \phi_3)$ where $(\phi_1^T,\phi_3^T)=(E(ZY_1)^T, E(ZY_2)^T)$ and $\phi_2=E(Zx^T).$ Then the identified set for $\theta$ is given by $\Theta(\phi)=\{\theta\in\Theta: \phi_1\leq \phi_2\theta\leq \phi_3\}.$ Suppose $\phi_2^{-1}$ exists. The support function for   $\Theta(\phi)$ is given by (denote $(x)_i$ as the $i$th component of $x$)\footnote{See Appendix \ref{s_appendix_support_function_interval_regression} in the supplementary material for detailed derivations of the support function in this example. Similar results but in a slightly different form are  presented in Bontemps et al. (2012).}:
$$
S_{\phi}(\nu)=\nu^T\phi_2^{-1}\left(\frac{\phi_1+\phi_3}{2}\right)+\alpha_{\nu}^T\left(\frac{\phi_3-\phi_1}{2}\right),\quad \nu\in\mathbb{S}^d
$$
where  $\alpha_{\nu}=( |(\nu^T\phi_2^{-1})_1|,..., |(\nu^T\phi_2^{-1})_d|)^T.$

 $\square$
\end{exm}

\begin{exm}[Missing data]\label{ex2.3}
  Consider a bivariate random vector $(Y,M)$ where $M$ is a binary random variable which takes the value $M=0$ when $Y$ is missing and $1$ otherwise. Here   $Y$ represents whether a treatment is successful ($Y=1$) or not ($Y=0$).  The parameter of interest is the probability  $\theta=P(Y=1)$.  This problem without the missing-at-random assumption has been extensively studied in the literature, see for example, Manski  and Tamer (2002), Manski (2003), etc.  Since $P(Y=1|M=0)$ cannot be recovered from the data,  the empirical evidence partially identifies $\theta$ and $\theta$ is characterized by the following moment restrictions:
    \begin{displaymath}
      P(Y=1|M=1)P(M=1) \leq \theta \leq  P(Y=1|M=1)P(M=1)+ P(M=0).
    \end{displaymath}
 Here, $\phi=(P(M=1), P(Y=1|M=1))=(\phi_1,\phi_2)$. The identified set is $\Theta(\phi)=[\phi_1\phi_2,\phi_1\phi_2+1-\phi_1]$,
 and its support function is:
$S_{\phi}(1) = \phi_1\phi_2+1-\phi_1$, $S_{\phi}(-1)=-\phi_1\phi_2$.

\end{exm}

\subsection{Nonparametric prior scheme for $(\phi, F)$}\label{sss_nonparametric_prior}


When the model only specifies a set of moment inequalities, we can place a non-parametric prior on the likelihood function through $F$, e.g., a Dirichlet process prior.   Since $\phi$ is point identified, we assume it can be rewritten as a measurable function of $F$ as  $\phi = \phi(F)$.  The prior distribution for $\phi$ is then deduced   from that of $F$ via $\phi(F)$. The Bayesian experiment is (we use the notation ``$\sim$" to mean ``distributed as")
$$
      X|F \sim F,\qquad F\sim \pi(F), \qquad \theta|\phi \sim \pi(\theta|\phi(F))
$$
For instance, in the interval censored data example \ref{ex2.1},  let $F$ be the joint CDF of $(Y_1, Y_2)$, then $(\phi_1, \phi_2)=\phi(F)=(E(Y_1|F), E(Y_2|F))$, and the identified set is modeled as  $\Theta(\phi)=[\phi_1(F), \phi_2(F)]$, which is a set-valued function of $F$.

The prior distribution $\pi(F)$ is a  distribution on $\mathcal{F}$. Examples of such a prior  include Dirichlet process priors (Ferguson 1973) and Polya tree (Lavine 1992). The case where $\pi(F)$ is a Dirichlet process prior in partially identified models is  proposed by \cite{FlorensSimoni2011}.


Let $p(F|D_n)$ denote the marginal posterior of $F$ which, by abuse of notation, can be written $p(F|D_n)\propto\pi(F)\prod_{i=1}^nF(X_i)$. The posterior distributions of $\phi$,    $\Theta(\phi)$, and the support function $S_{\phi}(\cdot)$ are deduced from the posterior of $F$, but do not depend on the prior on $\theta$. Moreover, it can be shown that $p(\theta|\phi(F), D_n)=\pi(\theta|\phi(F)).$ Then, for any measurable set $B\subset\Theta$, the marginal posterior probability of $\theta$ is given by, averaging over $F$:
    \begin{eqnarray*}\label{eq2.3}
      P(\theta\in B|D_n) & = & \int_{\mathcal{F}}P(\theta\in B |\phi(F), D_n)p(F|D_n)dF\nonumber\\
      & = & \int_{\mathcal{F}}\pi(\theta\in B|\phi(F))p(F|D_n)dF = E[\pi(\theta\in B|\phi(F))|D_n]
    \end{eqnarray*}
where the conditional expectation is taken with respect to the posterior of $F$. The above posterior is easy to calculate via simulation when $F$ has a Dirichlet process prior. 

An alternative prior scheme for $(\phi, F)$ consists in putting a prior   on $\phi$ directly. This  is particularly useful when there is informative prior information for $\phi$. It models the  unknown likelihood function semi-parametrically through reformulating $F$ as $F=F_{\phi, \eta}$ where $\eta$ is an infinite-dimensional nuisance parameter (often a density function) that is \textit{a priori} independent of $\phi$. The prior on $(\phi, F)$ is then deduced from the prior on $(\phi,\eta)$.  We describe this alternative semi-parametric prior in Appendix \ref{sss_semiparametric_prior}.


\section{Bayesian Inference for $\theta$}\label{s_nonparametric_prior}

\subsection{Putting priors on partially identified $\theta$}\label{app_prior_theta}

In this section we briefly discuss the meaning of the prior $\pi(\theta|\phi)$. As  stated in  Tamer (2010): ``(Partial identification) links conclusions drawn from various
empirical models to sets of assumptions made in a transparent way.
It allows researchers to examine the informational content of their
assumptions and their impacts on the inferences made.''

By imposing a  prior  on the partially identified parameter $\theta$, we reflect how prior beliefs and/or assumptions can impact the associated statistical inference. To illustrate the rationale of imposing such a prior, let us consider the missing data example (Example \ref{ex2.3}).    Writing $\alpha=P(Y=1|M=0)$, we then link $\theta$ with $\alpha$ by
\begin{equation}\label{eq2.2}
\theta=\phi_2\phi_1+\alpha(1-\phi_1).
\end{equation}
As $\phi$ is point identified, statistical inferences about $\theta$ therefore relies on the  treatment of $\alpha$. On the other hand, various ways of dealing with $\alpha$ reflect various researchers' prior beliefs, which also correspond to the ``informational content of their assumptions".

From a Bayesian point of view, this is fulfilled by putting a distribution on $\alpha$, as a prior $\pi(\alpha)$ supported on $[0,1]$ (possibly also depending on $\phi$).  The traditional exogeneity assumption such as missing-at-random, in this case, corresponds to a point mass prior on $\alpha=\phi_2=P(Y=1|M=1)$.   The more concentrating is the prior, the stronger are the assumptions we impose on the missing mechanism.  Such a prior distribution can also come from a previous study based on a different dataset that contain information about $\alpha$ where only summarizing statistics are available instead of the complete data set.  Above all, when no informative knowledge about $\alpha$ is available, a uniform prior on $[0,1]$ is imposed for $\alpha$, which reduces to Manski's bounds approach.

Given the imposed distribution $\pi(\alpha)$ that reflects researchers' assumptions or beliefs about the missing mechanism, we can deduce a conditional prior for $\theta$ through (\ref{eq2.2}) given $\phi=(\phi_1,\phi_2)$.
As a result, putting a prior on the partially identified parameter can be viewed as a way of incorporating researchers' assumptions on the missing mechanisms. This varies from the  traditional exogeneity approach to the  most  robust bounds approach, which also bridges point identification and partial identification.

\subsection{Posterior Consistency for $\theta$}
The shape of the posterior of a partially identified parameter    still relies upon its prior distribution asymptotically, which distinguishes from the asymptotic posterior behavior in the classical point identified case.  On the other hand, the support of the prior distribution of $\theta$  is revised after data are observed and eventually converges towards the true identified set asymptotically. The latter corresponds to the frequentist consistency of the posterior distribution for partially identified parameters. Posterior consistency  is one of the benchmarks  of a Bayesian procedure under consideration, which ensures that with a sufficiently large amount of data, it is nearly possible to discover the true identified set.

We assume there is a true value of $\phi$, denoted by $\phi_0$, which induces a true identified set $\Theta(\phi_0)$, and a true $F$, denoted by $F_{0}$. Our goal is to achieve the frequentist \textit{posterior consistency} for the partially identified parameter:   for any $\epsilon > 0$ there is $\tau \in (0,1]$ such that
$$
P(\theta\in\Theta(\phi_0)^{\epsilon}|D_n)\rightarrow^p 1,  \text{ and }\quad P(\theta\in\Theta(\phi_0)^{-\epsilon}|D_n)\rightarrow^p (1 - \tau).
$$
Here  $   \Theta(\phi)^{\epsilon} $ and $ \Theta(\phi)^{-\epsilon} $ are the $\epsilon$-envelope and $\epsilon$-contraction of $\Theta(\phi)$, respectively:
    \begin{equation}\label{eq_e_envelope}
      \Theta(\phi)^{\epsilon} = \{\theta\in\Theta: d(\theta, \Theta(\phi))\leq\epsilon\},\quad  \Theta(\phi)^{-\epsilon} = \{\theta\in\Theta(\phi): d(\theta, \Theta\backslash\Theta(\phi))\geq\epsilon\},
    \end{equation}
  with $\Theta\backslash\Theta(\phi) =\{\theta\in\Theta; \theta\notin \Theta(\phi)\}$ and $d(\theta,\Theta(\phi))=\inf_{x\in\Theta(\phi)}\|\theta-x\|$. Note that this result still carries over  when  $\theta$ is point identified, in which  case $\Theta(\phi)^{\epsilon}$ is an $\epsilon$-ball around $\theta$, $\Theta(\phi)^{-\epsilon}$ is empty, and $\tau=1.$

The likelihood function is endowed with a prior through either the nonparametric prior $\pi(F)$ as described in Section \ref{sss_nonparametric_prior} or the semi-parametric prior $\pi(\phi)$ as described in Appendix \ref{sss_semiparametric_prior}. We assume that the priors $\pi(F)$ and $\pi(\phi)$ specified for $F$ and $\phi$ are such that the corresponding posterior distribution of $p(\phi|D_n)$ is consistent.

\begin{assum}\label{ass3.1}
  At least one of the following holds:

    \begin{itemize}
      \item[(i)] The measurable function $\phi(F):\mathcal{F}\rightarrow\Phi$ is continuous. The prior $\pi(F)$ is such that the posterior $p(F|D_{n})$ satisfies:

        \begin{displaymath}
          \int_{\mathcal{F}} m(F)p(F|D_n)dF \rightarrow^p  \int_{\mathcal{F}} m(F)\delta_{F_0}(dF)
        \end{displaymath}

        for any bounded and continuous function $m(\cdot)$ on $\mathcal{F}$ where $\delta_{F_0}$ is the Dirac function at the true distribution function $F_0$;

      \item[(ii)] The prior $\pi(\phi)$ is such that the posterior $p(\phi|D_{n})$ satisfies:

        \begin{displaymath}
          \int_{\Phi} m(\phi)p(\phi|D_n)d\phi \rightarrow^p  \int_{\Phi} m(\phi)\delta_{\phi_0}(d\phi)
        \end{displaymath}

        for any bounded and continuous function $m(\cdot)$ on $\Phi$ where $\delta_{\phi_0}$ is the Dirac function at the true  $\phi_0$.

    \end{itemize}
\end{assum}

Assumptions \ref{ass3.1} \textit{(i)} and \textit{(ii)} refer to the nonparametric and semi-parametric prior scheme respectively, and are verified by many nonparametric and semi-parametric priors. Examples are: Dirichlet process priors, Polya Tree process priors, Gaussian process priors, etc.  For instance, when $\pi(F)$ is the Dirichlet process prior, the second part of Assumption \ref{ass3.1} \textit{(i)} was proved in Ghosh and Ramamoorthi (2003, Theorem 3.2.7) while the condition that $\phi(F)$ is continuous in $F$ is verified in many examples relevant for applications. For instance, in Example \ref{ex2.1}, $\phi(F)=(E(Y_1|F),E(Y_2|F))^T$ and in Example \ref{ex2.2}, $\phi(F)=(E(ZY_1|F), E(ZX^T|F), E(ZY_2|F))$, which are all bounded linear functionals of $F$. 
We refer to \cite{GhoshRamamoorthi2003} for examples and sufficient conditions for this assumption.

\begin{assum}[Prior for $\phi$]\label{ass3.2} For any $\epsilon>0$ there are  measurable sets $A_2\subset A_1\subset\Phi$ such that  $0<\pi(\phi\in A_i)\leq 1$,  $i=1,2$ and \\
(i)  for all $\phi\in A_1$, $\Theta(\phi_0)^{\epsilon}\cap \Theta(\phi)\neq\emptyset$; for all $\phi\notin A_1$, $\Theta(\phi_0)^{\epsilon}\cap \Theta(\phi)=\emptyset$,\\
(ii) for all $\phi\in A_2$, $\Theta(\phi_0)^{-\epsilon}\cap \Theta(\phi)\neq\emptyset$; for all $\phi\notin A_2$, $\Theta(\phi_0)^{-\epsilon}\cap \Theta(\phi)=\emptyset$.
\end{assum}



Assumption \ref{ass3.2} is satisfied as long as the identified set $\Theta(\phi)$ is bounded and the prior of $\phi$ spreads over a large support of the parameter space. This assumption allows us to prove the posterior consistency without assuming the prior $\pi(\theta|\phi)$ to be a continuous function of $\phi$, and therefore priors like $I_{\phi_1<\theta<\phi_2}$ in the interval censoring data example are allowed. Under this assumption 
the conditional prior probability of the $\epsilon$-envelope of the true identified set can be approximated by a continuous function, that is, there is a sequence of bounded and continuous functions $h_m(\phi)$ such that (see lemma \ref{LemB.1} in the appendix) almost surely in $\phi$:
$$
\pi (\theta\in\Theta(\phi_0)^{\epsilon}|\phi) = \lim_{m\rightarrow\infty}h_m(\phi).
$$
\noindent A similar approximation holds for the conditional prior of the $\epsilon$-contraction $\pi(\theta\in\Theta(\phi_0)^{-\epsilon}|\phi)$.

\begin{assum}[Prior for $\theta$]\label{ass3.3}
For any $\epsilon>0$, and $\phi\in\Phi$,  $\pi(\theta\in\Theta(\phi)^{-\epsilon}|\phi)<1.$
\end{assum}

 In the special case when $\theta$ is point identified ($\Theta(\phi)$ is a singleton), the $\epsilon$-contraction  is empty and thus $\pi(\theta\in\Theta(\phi)^{-\epsilon}|\phi)=0.$

Assumption \ref{ass3.3}  is an assumption on the prior for $\theta$, which means the identified set should be \textit{sharp} with respect to the prior information. Roughly speaking, the support of the prior should not be a proper subset of any $\epsilon$-contraction of the identified set $\Theta(\phi)$. If otherwise the prior information restricts $\theta$ to be inside a strict subset of $\Theta(\phi)$ so that Assumption \ref{ass3.3} is violated, then that prior information should be taken into account in order to shrink $\Theta(\phi)$ to a sharper set. In that case, the posterior will asymptotically concentrate around a set that is smaller than the   set identified by the data alone. Remark that assumption \ref{ass3.3} is not needed for the first part of Theorem \ref{th3.1} below.

 The following theorem gives the posterior consistency for partially identified parameters. 
\begin{thm}\label{th3.1}
  Under Assumptions \ref{ass3.1} and \ref{ass3.2}, for any $\epsilon>0$,

    \begin{displaymath}
      P(\theta\in \Theta(\phi_{0})^{\epsilon}|D_n)\rightarrow^p 1.    \end{displaymath}
      If Assumption \ref{ass3.3} is further satisfied, then  there is $  \tau \in (0,1]$ such that
      $$
  P(\theta\in\Theta(\phi_0)^{-\epsilon}|D_n)\rightarrow^p (1 - \tau).
      $$
\end{thm}

\section{Bayesian Inference of Support Function}\label{ssf}
Our analysis focuses on identified sets which are closed and convex. These sets are completely determined by their support functions, and efficient estimation of support functions may lead to optimality of estimation and inference of the identified set. As a result, much of the new development in the partially identified literature focuses on the support function, e.g., Kaido and Santos (2013),  Kaido (2012),  Beresteanu and Molinari (2008), Bontemps et al. (2012). 

This section develops Bayesian analysis for the support function $S_{\phi}(\nu)$ of the identified set $\Theta(\phi)$. We consider a more specific partially identified model: the \textit{moment inequality model} which is described in section \ref{ss_5.1} below. Bayesian inference for the support function has two main interests. First, it provides an alternative way to characterize and perform estimation of the identified set $\Theta(\phi)$, which in many cases is relatively easy for computations and simulations. Second, it allows us to construct a two-sided BCS for $\Theta(\phi)$ that is also asymptotically equivalent to a frequentist confidence set. In this section we first develop a  local linearization in $\phi$ of the support function.  As the support function itself is ``point identified'', we prove that its  posterior satisfies the Bernstein-von Mises theorem. This result is \textit{per se} of particular interest in the nonparametric Bayesian literature.


\subsection{Moment Inequality Model}\label{ss_5.1}
The \textit{moment inequality model} assumes that $\theta$ satisfies $k$ moment restrictions:
\begin{equation}\label{eq4.1}
\Psi(\theta, \phi)\leq 0,\quad \Psi(\theta, \phi)=(\Psi_1(\theta,\phi),..., \Psi_k(\theta,\phi))^T
\end{equation}
where $\Psi: \Theta\times\Phi\rightarrow\mathbb{R}^k$ is a known function of $(\theta, \phi)$.  The identified set can be characterized as:
    \begin{equation}\label{eq4.2old}
      \Theta(\phi)=\{\theta\in\Theta: \Psi(\theta, \phi)\leq 0\}.
    \end{equation}
Since most of the partially identified models can be characterized as moment inequality models, model (\ref{eq4.1})-(\ref{eq4.2old}) has received extensive attention in the  literature.

We assume  each component of $\Psi(\theta,\phi)$ to be a convex function of $\theta$ for every $\phi\in\Phi$,  as stated in the next assumption.

\begin{assum}\label{ass5.1}
  $\Psi(\theta,\phi)$ is continuous in $(\theta,\phi)$ and convex in $\theta$ for every $\phi\in\Phi$.
\end{assum}

Let us consider the support function $S_{\phi}(\cdot):\mathbb{S}^{d}\rightarrow\mathbb{R}$ of the identified set $\Theta(\phi)$. We restrict its domain to the unit sphere $\mathbb{S}^{d}$ in $\mathbb{R}^{d}$ since $S_{\phi}(\nu)$ is positively homogeneous in $\nu$. Under Assumption \ref{ass5.1} the support function is the optimal value of an ordinary convex program:
\begin{displaymath}
  S_{\phi}(\nu) = \sup_{\theta\in\Theta}\{\nu^T\theta;\:  \Psi(\theta, \phi)\leq 0\}.
\end{displaymath}
Therefore, it also admits a Lagrangian representation (see Rockafellar 1970, chapter 28):
\begin{equation}\label{eq5.1}
  S_{\phi}(\nu) = \sup_{\theta\in\Theta}\{\nu^T\theta - \lambda(\nu,\phi)^T\Psi(\theta,\phi)\},
\end{equation}

\noindent where $\lambda(\nu,\phi): \mathbb{S}^{d}\times\mathbb{R}^{d_{\phi}}\rightarrow\mathbb{R}_{+}^{k}$ is a $k$-vector of Lagrange multipliers.\\
\indent We denote by $\Psi_{S}(\theta,\phi_{0})$ the $k_{S}$-subvector of $\Psi(\theta,\phi_{0})$ containing the constraints that are strictly convex functions of $\theta$ and by $\Psi_{L}(\theta,\phi_{0})$ the $k_{L}$ constraints that are linear in $\theta$.  So $k_S + k_L = k$. The corresponding Lagrange multipliers are denoted by $\lambda_{S}(\nu,\phi_{0})$ and $\lambda_{L}(\nu,\phi_{0})$, respectively, for $\nu\in\mathbb{S}^{d}$. Moreover, define $\Xi(\nu,\phi) = \arg\max_{\theta \in\Theta}\{\nu^T\theta; \: \Psi(\theta,\phi)\leq 0\}$  as the \textit{support set} of $\Theta(\phi)$. Then, by definition,
$$
\nu^T\theta=S_{\phi}(\nu),\quad \forall\theta\in\Xi(\nu,\phi).
$$
We also denote by $\nabla_{\phi}\Psi(\theta,\phi)$ the $k\times d_{\phi}$ matrix of partial derivatives of $\Psi$ with respect to $\phi$, and by $\nabla_{\theta}\Psi_{i}(\theta,\phi)$ the $d$-vector of partial derivatives of $\Psi_{i}$ with respect to $\theta$ for each $i\leq k$. In addition, let
$$Act(\theta,\phi)\equiv\{i\leq k;\: \Psi_i (\theta,\phi)=0\}$$
be the set of the inequality active constraint indices. For some $\delta>0$, let $B(\phi_{0}, \delta)=\{\phi\in\Phi; \: \|\phi - \phi_{0}\|\leq \delta\}$.

We assume the following:

\begin{assum}\label{ass5.2}
  The true value $\phi_{0}$ is in the interior of $\Phi$, and $\Theta$ is convex and compact.
\end{assum}

\begin{assum}\label{ass5.3} There is some $\delta>0$ such that for all $\phi \in B(\phi_{0}, \delta)$, we have:\\
(i)  the  matrix $\nabla_{\phi}\Psi(\theta,\phi)$ exists and is continuous in $(\theta,\phi)$;\\
(ii) the set $\Theta(\phi)$ is non empty;\\
(iii)  there exists a $\theta\in\Theta$ such that $\Psi(\theta,\phi)<0$;\\
(iv) $\Theta(\phi)$ belongs to the interior of $\Theta$;\\
(v) for every $i\in Act(\theta,\phi_{0})$, with $\theta\in\Theta(\phi_{0})$, the vector $\nabla_{\theta}\Psi_{i}(\theta,\phi)$ exists and is continuous in $(\theta,\phi)$ for every $\phi\in B(\phi_{0},\delta)$ and $\theta\in\Theta(\phi)$.
\end{assum}

Assumption \ref{ass5.3} \textit{(iii)} is the Slater's condition which is a sufficient condition for strong duality to hold. It implies Assumption \ref{ass5.3} \textit{(ii)}. However, we keep both conditions because in order to establish some technical results we only need condition \textit{(ii)} which is weaker.



The next assumption concerns the inequality active constraints. Assumption \ref{ass5.4} requires that the active inequality constraints gradients $\nabla_{\theta}\Psi_{i}(\theta,\phi_{0})$ be linearly independent. This assumption guarantees that a $\theta$ which solves the optimization problem (\ref{eq5.1}) with $\phi = \phi_{0}$ satisfies the Kuhn-Tucker conditions. Alternative assumptions that are weaker than Assumption \ref{ass5.4} could be used, but the advantage of Assumption \ref{ass5.4} is that it is easy to check. 

\begin{assum}\label{ass5.4}
   For any $  \theta\in\Theta(\phi_{0})$,  the gradient vectors $\{\nabla_{\theta}\Psi_{i}(\theta,\phi_{0})\}_{i\in Act(\theta,\phi_{0})}$ are linearly independent
\end{assum}

 The following assumption is key for our analysis, and is sufficient for the differentiability of the support function at $\phi_0$:
\begin{assum}\label{ass5.6}
At least one of the following holds:
    \begin{itemize}
      \item[\textit{(i)}] For the ball $B(\phi_0,\delta)$ in Assumption \ref{ass5.3},  for every $(\nu,\phi)\in\mathbb{S}^{d}\times B(\phi_{0},\delta)$, $\Xi(\nu,\phi)$ is a singleton;
      \item[\textit{(ii)}] There are linear constraints in $\Psi(\theta,\phi_0)$, which are also separable in $\theta$, that is, $k_L>0$ and $\Psi_{L}(\theta,\phi_{0}) = A_{1}\theta + A_{2}(\phi_{0})$ for some function $A_{2}: \Phi\rightarrow \mathbb{R}^{k_{L}}$ (not necessarily linear) and some $(k_L\times d)$-matrix $A_{1}$.
    \end{itemize}
\end{assum}

 Assumption \ref{ass5.6} is particularly important for the linearization of the support function that we develop in Section \ref{ss_5.2}. In fact, if one of the two parts of Assumption \ref{ass5.6} holds then the support function is differentiable at $\phi$ for every $(\nu,\phi)\in\mathbb{S}^{d}\times B(\phi_{0}, \delta)$,  and we have a closed form for its derivative. This assumption also plays one of the key roles in the  study of asymptotic efficiency by Kaido and Santos (2013).

The last set of assumptions   will be used to prove the Bernstein-von Mises theorem for $S_{\phi}(\cdot)$ and allows to strengthen the result of Theorem \ref{Lem5.2} below. The first three assumptions are (local) Lipschitz equi-continuity assumptions.

\begin{assum}\label{ass5.5}
  For the ball $B(\phi_{0},\delta)$ in Assumption \ref{ass5.3}, for some $K>0$ and $\forall \phi_{1},\phi_{2}\in B(\phi_{0},\delta)$:
    \begin{itemize}
    \item[(i)] $\sup_{\nu\in\mathbb{S}^d}\|\lambda(\nu,\phi_{1}) - \lambda(\nu,\phi_{2})\| \leq K\|\phi_{1} - \phi_{2}\|$;
    \item[(ii)] $\sup_{\theta\in\Theta}\|\nabla_{\phi}\Psi(\theta,\phi_{1}) - \nabla_{\phi}\Psi(\theta,\phi_{2})\| \leq K\|\phi_{1} - \phi_{2}\|$;
    \item[(iii)] $\|\nabla_{\phi}\Psi(\theta_1,\phi_{0}) - \nabla_{\phi}\Psi(\theta_2,\phi_{0})\| \leq K\|\theta_{1} - \theta_{2}\|$, for every $\theta_1, \theta_2\in \Theta$;
    \item[(iv)] If $\Xi(\nu,\phi_0)$ is a singleton for any $\nu$ in  some compact subset $W\subseteq \mathbb{S}^d$, and if the correspondence $(\nu,\phi) \mapsto \Xi(\nu,\phi)$ is upper hemicontinous on $\mathbb{S}^d\times B(\phi_0,\delta)$ then there exists $\varepsilon =O(\delta)$ such that $\Xi(\nu,\phi_1) \subseteq \Xi^{\varepsilon}(\nu,\phi_0)$.
  \end{itemize}
\end{assum}
Here $\|\nabla_{\phi}\Psi(\theta,\phi)\|$ denotes the Frobenius norm of the matrix. The above conditions are not stringent.  In particular,  condition \textit{(iv)} is easy to understand when $\Xi(p,\phi)$ is a singleton, that is, when the optimization problem for the support function has a unique solution, for each $\phi\in B(\phi_{0},\delta)$. Then $\Xi(\nu,\phi_1)$ and $\Xi(\nu,\phi_0)$ are singletons that are close to each other, and $\Xi(\nu,\phi_0)^{\varepsilon}$ is a small ball around $\Xi(\nu,\phi_0)$.

\indent We show in the following example that Assumptions \ref{ass5.1}-\ref{ass5.5}  are easily satisfied.

\begin{exm}[Interval censored data - \textit{continued}]
  The setup is the same as in Example \ref{ex2.1}. Assumption \ref{ass5.2} is verified if $Y_{1}$ and $Y_{2}$ are two random variables with finite first moments $\phi_{0,1}$ and $\phi_{0,2}$, respectively. Moreover, $\Psi(\theta,\phi) = (\phi_{1} - \theta, \theta - \phi_{2})^T$, $\phi = (\phi_{1},\phi_{2})^T$,
\begin{displaymath}
  \nabla_{\phi}\Psi(\theta,\phi) = \left(\begin{array}
  {cc} 1 & 0\\ 0 & -1
\end{array}\right)
\end{displaymath}
so that Assumptions \ref{ass5.1}, \ref{ass5.2} and \ref{ass5.3} \textit{(i)}-\textit{(ii)} are trivially satisfied. Assumption \ref{ass5.3} \textit{(iii)} holds for every $\theta$ inside $(\phi_{1},\phi_{2})$; Assumption \ref{ass5.3} \textit{(iv)} is satisfied if $\phi_{1}$ and $\phi_{2}$ are bounded. To see that Assumptions \ref{ass5.3} \textit{(v)} and \ref{ass5.4} are satisfied note that $\forall \theta < \phi_{0,1}$ we have $Act(\theta,\phi_{0}) = \{1\}$, $\forall \theta > \phi_{0,2}$ we have $Act(\theta,\phi_{0}) = \{2\}$ while $\forall \theta \in[\phi_{0,1}, \phi_{0,2}]$ we have $Act(\theta,\phi_{0}) = \emptyset$. Assumption \ref{ass5.6} \textit{(i)} and \textit{(ii)} are both satisfied since the support set takes the values $\Xi(1,\phi)=\phi_2$ and $\Xi(-1,\phi)=-\phi_1$ and the constraints in $\Psi(\theta,\phi_{0})$ are both linear with $A_{1} = (-1,1)^T$ and $A_{2}(\phi_{0}) = \nabla_{\phi}\Psi(\theta,\phi_{0})\phi_{0}$.\\
\indent  Assumptions \ref{ass5.5} \textit{(ii)}-\textit{(iii)} are naturally satisfied because $\nabla_{\phi} \Phi(\theta, \phi)$ does not depend on $(\theta, \phi)$.  The Lagrange multiplier is $\lambda(\nu,\phi) = (-\nu I(\nu<0), \nu I(\nu\geq 0))^{T}$ so that Assumption \ref{ass5.5} \textit{(i)} is satisfied since the norm is equal to $0$. Finally, the support set $\Xi(\nu,\phi) = \phi_{1}I(\nu<0) + \phi_{2}I(\nu\geq 0)$ is a singleton for every $\phi\in B(\phi_{0}, \delta)$ and $\Xi(\nu,\phi_{0})^{\varepsilon} = \{\theta\in\Theta;\|\theta - \theta_{*}\| \leq \varepsilon \}$ where $\theta_{*} = \Xi(\nu,\phi_{0}) = \phi_{0,1}I(\nu<0) + \phi_{0,2}I(\nu\geq 0)$. Therefore, $\|\Xi(\nu,\phi) - \theta_{*}\| \leq \delta$ and Assumption \ref{ass5.5} \textit{(iv)} holds with $\varepsilon = \delta$. $\square$
\end{exm}

\subsection{Asymptotic Analysis}\label{ss_5.2}
The support function of the closed and convex set $\Theta(\phi)$ 
admits directional derivatives in $\phi$, see e.g. \cite{MilgromSegal2002}. Moreover, if Assumption \ref{ass5.6} holds for a particular value $(\nu,\phi)$, then $S_{\phi}(\nu)$ is differentiable at $\phi$ and its derivative is equal to the left and right directional derivatives. The next theorem exploits this fact  and states that the support function can be locally approximated by a linear function of $\phi$.   

\begin{thm}\label{Lem5.2}
  If Assumptions \ref{ass5.1}-\ref{ass5.6} hold with $\delta = r_{n}$ for some $r_n=o(1)$,  then there is  $N\in\mathbb{N}$ such that for every $n\geq N$,  there exist: (i) a real function $f(\phi_{1},\phi_{2})$ defined for every $\phi_1, \phi_2\in B(\phi_{0},r_{n})$,  
  (ii) a Lagrange multiplier  function $\lambda(\nu,\phi_0): \mathbb{S}^{d}\times\mathbb{R}^{d_{\phi}}\rightarrow\mathbb{R}_{+}^{k}$, and (iii)   a Borel measurable mapping  $\theta_{*}(\nu):\mathbb{S}^{d}\rightarrow \Theta$  satisfying $\theta_{*}(\nu)\in\Xi(\nu,\phi_{0})$  for all $\nu\in\mathbb{S}^{d}$,  such that for every $\phi_1, \phi_2\in B(\phi_{0},r_{n})$:

  \begin{displaymath}
    \sup_{\nu\in\mathbb{S}^{d}}\left|\left(S_{\phi_1}(\nu) - S_{\phi_2}(\nu)\right) - \lambda(\nu,\phi_{0})^T\nabla_{\phi}\Psi(\theta_{*}(\nu),\phi_{0})[\phi_1-\phi_2]\right| = f(\phi_{1},\phi_{2})
  \end{displaymath}

  \noindent and $\frac{f(\phi_{1},\phi_{2})}{\|\phi_{1} - \phi_{2}\|} \rightarrow 0$ uniformly in $\phi_1, \phi_2\in B(\phi_{0},r_{n})$ as $n\rightarrow \infty$.
\end{thm}

We remark that the functions $\lambda$ and $\theta_*$ do not depend on the specific choice of $\phi_1$ and $\phi_2$ inside $B(\phi_0,r_n)$, but only on $\nu$ and the true value $\phi_0$.   The expansion can also be viewed as  stochastic when $\phi_1, \phi_2$ are interpreted as random variables associated with the posterior distribution $P(\phi|D_{n})$. This interpretation  is particularly useful to understand Theorems \ref{th5.2} and \ref{th5.3}.

With the approximation given in the theorem we are now ready to state posterior consistency (with concentration rate) and asymptotic normality of the posterior distribution of $S_{\phi}(\nu)$. The posterior consistency of the support function is also based upon the posterior concentration rate for $\phi$. In a semi-parametric Bayesian model where $\phi$ is point identified, the posterior of $\phi$ achieves a near-parametric concentration rate under proper prior conditions. Since our goal is to study the posterior of $S_{\phi}(\nu)$, we state a high-level assumption on the posterior of $\phi$ as follows, instead of deriving it from more general conditions.

\begin{assum}\label{ass4.1}The marginal posterior of $\phi$ is such that, for some $C>0$,
$$P(\|\phi-\phi_0\|\leq Cn^{-1/2}(\log n)^{1/2}|D_n)\rightarrow^p1.$$
\end{assum}

This assumption is 
a standard result in semi/non-parametric Bayesian literature. If we place a nonparametric prior on $F$ as described in Section \ref{sss_nonparametric_prior}, the notation used in this assumption is a shorthand for $$P(\|\phi(F)-\phi(F_0)\|\leq Cn^{-1/2}(\log n)^{1/2}|D_n)\rightarrow^p1.$$ When the likelihood function is unknown, a formal derivation of this assumption for a semi-parametric prior of $(\phi, F)$ will be presented in Appendix \ref{s_appendix_posterior_concentration_phi}. 

The next theorem gives the contraction rate for the posterior of the support function.

\begin{thm}\label{th5.2}
  Under Assumption \ref{ass4.1} and the Assumptions of Theorem \ref{Lem5.2} with $r_n=\sqrt{(\log n)/n}$, for some $C>0$,
    \begin{equation}\label{eq5.2}
      P\left(\left.\sup_{\nu\in\mathbb{S}^{d}}|S_{\phi}(\nu) - S_{\phi_{0}}(\nu)|<C(\log n)^{1/2}n^{-1/2} \right|D_{n}\right) \rightarrow^{p} 1.
    \end{equation}
\end{thm}

\begin{remark}
The above result holds for both nonparametric and semi-parametric prior on $(\phi,F)$. The concentration rate, as given in the theorem, is nearly  parametric: $\sqrt{\frac{\log n}{n}}$ and is the same as the rate in assumption \ref{ass4.1}. Thus, when the posterior for $\phi$ contracts at the rate $n^{-1/2}$,  the same holds for the posterior of the support function. The posterior probability in the theorem is now converging to zero, instead of being smaller than an arbitrarily small constant. This often gives rise to the term $\sqrt{\log n}$, which  arises commonly in the posterior concentration rate literature (e.g., Ghosal et al. 2000, Shen and Wasserman 2001). The same rate of convergence in the frequentist perspective has been achieved by Chernozhukov et al. (2007),  Beresteanu and Molinari  (2008),  Kaido and Santos (2013), among others, when estimating the identified set.
\end{remark}

We now state a Bernstein-von Mises (BvM) theorem for the support function. This theorem is valid under the assumption that a BvM  theorem holds for the posterior distribution of the   identified parameter $\phi$. We denote by $\|\cdot\|_{TV}$ the total variation distance, that is, for two probability measures $P$ and $Q$, $$\|P-Q\|_{TV} \equiv \sup_{B}|P(B) - Q(B)|$$ where $B$ is an element of the $\sigma$-algebra on which $P$ and $Q$ are defined.

\begin{assum}\label{ass5.9}
  Let $P_{\sqrt{n}(\phi - \phi_{0})|D_{n}}$ denote the posterior distribution of $\sqrt{n}(\phi - \phi_{0})$. We assume
    \begin{displaymath}
      \|P_{\sqrt{n}(\phi - \phi_{0})|D_{n}} - \mathcal{N}({\Delta}_{n,\phi_{0}}, I_0^{-1})\|_{TV} \rightarrow^p 0
    \end{displaymath}

  \noindent where $\mathcal{N}$ denotes the $d_{\phi}$-dimensional normal distribution, ${\Delta}_{n,\phi_{0}}\equiv n^{-1/2}\sum_{i=1}^{n}I_0^{-1}l_{\phi_{0}}(X_{i})$, $l_{\phi_{0}}$ is the semi-parametric efficient score function of the model and $I_0$ denotes the semi-parametric efficient  information matrix.
\end{assum}

As we focus on the partial identification features, we state the above assumption as a high-level condition instead of proving it.  We refer to Bickel and Kleijn (2012) and  Rivoirard and Rousseau (2012) for primitive conditions of  this assumption in semi-parametric models. Remark that $l_{\phi_{0}}$ and $I_0$ also depend  on the true  likelihood function. The \textit{semi-parametric efficient score function} and the \textit{semi-parametric efficient information} contribute to the stochastic local asymptotic normality (LAN, Le Cam 1986) expansion of the integrated likelihood, which is necessary in order to get the BvM result in Assumption \ref{ass5.9}. A precise definition of $l_{\phi_{0}}$ and $I_0$ may be found in \cite{vanderVaart2002} (Definition 2.15). In particular, they become the usual score function (first derivative of the likelihood) and Fisher's information matrix when the true likelihood is fully parametric.

Below we denote $P_{\sqrt{n}(S_{\phi}(\nu) - S_{\phi_{0}}(\nu))|D_{n}} $ as the posterior distribution of $\sqrt{n}(S_{\phi}(\nu) - S_{\phi_{0}}(\nu)).$
\begin{thm}\label{th5.3}
  Let Assumption \ref{ass5.9} hold. If the assumptions of Theorem \ref{th5.2} and Assumption \ref{ass5.5} hold with $\delta = r_{n} = \sqrt{(\log n)/n}$, then for any $\nu\in\mathbb{S}^{d}$,
    \begin{displaymath}
      \|P_{\sqrt{n}(S_{\phi}(\nu) - S_{\phi_{0}}(\nu))|D_{n}} - \mathcal{N}(\bar{\Delta}_{n,\phi_{0}}(\nu), \bar{I}_{0}^{-1}(\nu))\|_{TV} \rightarrow^p 0,
    \end{displaymath}
  \noindent where $\bar{\Delta}_{n,\phi_{0}}(\nu)= \lambda(\nu,\phi_{0})^T\nabla_{\phi}\Psi(\theta_{*}(\nu),\phi_{0}){\Delta}_{n,\phi_{0}}$ and $$\bar{I}_{0}^{-1}(\nu) = \lambda(\nu,\phi_{0})^T\nabla_{\phi}\Psi(\theta_{*}(\nu),\phi_{0}) I_0^{-1} \nabla_{\phi}\Psi(\theta_{*}(\nu),\phi_{0})^{T}\lambda(\nu,\phi_{0}).$$
\end{thm}

 The asymptotic mean and covariance matrix can be easily estimated by replacing $\phi_{0}$ by any consistent estimator $\hat{\phi}$. Thus, $\theta_{*}(\nu)$ will be replaced by an element $\hat{\theta}_{*}(\nu)\in\Xi(\nu,\hat{\phi})$ and an estimate of $\lambda(\nu,\phi_{0})$ will be obtained by numerically solving the ordinary convex program in (\ref{eq5.1}) with $\phi_{0}$ replaced by $\hat{\phi}$.

\begin{remark} The posterior asymptotic variance of the support function $\bar{I}_{0}^{-1}$ is the same as that of the frequentist estimator obtained by Kaido and Santos (2013, Theorem 3.2). Both are derived based on a linear expansion of the support function.  This implies that the  Bayesian estimation of the  support function   is also asymptotically semi-parametric efficient in the frequentist sense. On the other hand,  there is also a major difference between our results and theirs because when studying the posterior distributions, we do not have an empirical process as Kaido and Santos (2013) do. This requires us to develop a different strategy to prove the linear expansion given in Theorem \ref{Lem5.2} as well as the asymptotic normalities given in Theorems  \ref{th5.3} and \ref{t5.4} below.  This also achieves a more strengthened result because the expansion   in Theorem \ref{Lem5.2} is  uniformly valid in a neighborhood of $\phi_0$.   


\end{remark}


\begin{remark}
The support function $S_{\phi}(\cdot)$ is a stochastic process with realizations in   $\mathcal{C}(\mathbb{S}^{d})$, the space of bounded continuous functions on $\mathbb{S}^{d}$.  Despite of the pointwise convergence in Theorem  \ref{th5.3} for each fixed $\nu$, however, the posterior distribution of the process $\sqrt{n}(S_{\phi}(\cdot) - S_{\phi_{0}}(\cdot))$ does not converge to a Gaussian measure on  $\mathcal{C}(\mathbb{S}^{d})$ in the total variation distance. Roughly speaking, the convergence in total variation would require the existence of a Gaussian measure $\mathbb{G}(\cdot)$ on $\mathcal{C}(\mathbb{S}^{d})$ such that uniformly in all Borel measurable sets $B$ of $\mathcal{C}(\mathbb{S}^{d})$,
  \begin{equation}\label{e5.3}
    |P_{\sqrt{n}(S_{\phi}(\cdot) - S_{\phi_{0}}(\cdot))|D_{n}}(B) - \mathbb{G}(B)| \rightarrow^{p} 0,
  \end{equation}
where $P_{\sqrt{n}(S_{\phi}(\cdot) - S_{\phi_{0}}(\cdot))|D_{n}}$ denotes the posterior distribution of the centered support function. However, in general (\ref{e5.3}) does not hold   uniformly in all the Borel sets $B$. Such a negative result can be made rigorous, and is generally known, see e.g., Freedman (1999) or Leahu (2011).
$\square$
 \end{remark}

On the positive side,  a \emph{weak} Bernstein-von Mises theorem holds  with respect to the weak topology. More precisely, let $\mathbb{G}$ be a Gaussian measure on $\mathcal{C}(\mathbb{S}^{d})$ with mean function $\bar{\Delta}_{n,\phi_{0}}(\cdot)= \lambda(\cdot,\phi_{0})^T\nabla_{\phi}\Psi(\theta_{*}(\cdot),\phi_{0}){\Delta}_{n,\phi_{0}}$ and covariance operator with kernel
  $$\bar{I}_{0}^{-1}(\nu_{1},\nu_{2}) = \lambda(\nu_{1},\phi_{0})^T\nabla_{\phi}\Psi(\theta_{*}(\nu_{1}),\phi_{0}) I_0^{-1} \nabla_{\phi}\Psi(\theta_{*}(\nu_{2}),\phi_{0})^{T}\lambda(\nu_{2},\phi_{0}),\qquad \forall \nu_{1},\nu_{2}\in\mathbb{S}^{d}.$$  We then have the following theorem. For a set $B$,  denote by $\partial B$ the boundary set of $B$.

  \begin{thm}\label{t5.4} Let $\mathcal{B}$ be the class of Borel measurable sets in $\mathcal{C}(\mathbb{S}^{d})$ such that $\mathbb{G}(\partial B)=0$.   Under the assumptions of Theorem \ref{th5.3},
    \begin{equation}\label{eq_weak_convergence_posterior_equivalent}
      \sup_{B\in\mathcal{B}}\left|P_{\sqrt{n}(S_{\phi}(\cdot) - S_{\phi_{0}}(\cdot))|D_{n}}(B) - \mathbb{G}(B)\right| \rightarrow^{p} 0.
    \end{equation}
    Let  `$\Rightarrow$' denote weak convergence on the class of probability measures on $\mathcal{C}(\mathbb{S}^{d}).$ 
    Then equivalently,
        \begin{equation}\label{eq_weak_convergence_posterior}
      P_{\sqrt{n}(S_{\phi}(\cdot) - S_{\phi_{0}}(\cdot))|D_{n}} \Rightarrow \mathbb{G}(\cdot).
    \end{equation}
  \end{thm}

\subsection{Models with moment equalities}\label{ss_5.3}
Our  analysis carries over when the model contains both moment equalities and inequalities if the moment equality functions are affine functions. This case is more general than the previous one. Suppose that the identified set writes as

    \begin{eqnarray}\label{eq_5.3}
      \Theta(\phi) = \{\theta\in\Theta; && \Psi_{i}(\theta,\phi)\leq 0,\: i=1,\ldots,k_1 \quad\textrm{and }\nonumber\\
      && a_{i}^{T}\theta + b_{i}(\phi) = 0, \: i = k_1 + 1,\ldots, k_1 + k_2\}
    \end{eqnarray}

\noindent where $a_{i}$ is a $(d\times 1)$-vector and $b_{i}$ is a continuous real-valued function of $\phi$ for all $i$. Let $k_1$ denote the number of moment inequalities, $k_2$ denote the number of moment equalities, and  $k=k_1 + k_2$. We then define $\Psi(\theta,\phi)$ as the $(k\times 1)$ vector whose first $k_1$ components are the functions $\Psi_{i}(\theta,\phi)$, $i=1,\ldots,k_1$ and the last $k_2$ components are the functions $a_{i}^{T}\theta + b_{i}(\phi)$, $i=k_1 + 1, \ldots, k_1 + k_2$.  

  The set $\Theta(\phi)$ is closed and convex (with an empty interior) since it is the intersection of a closed and convex set with closed hyperplanes. In this case, the support function   still has a  Lagrangian representation  as:
    \begin{displaymath}
      S_{\phi}(\nu) = \sup_{\theta\in\Theta}\{\nu^{T}\theta - \lambda(\nu,\phi)^{T}\Psi(\theta,\phi)\},
    \end{displaymath}

\noindent where $\lambda(\nu,\phi):\mathbb{S}^{d}\times\mathbb{R}^{d_{\phi}}\rightarrow \mathbb{R}_{+}^{k_1}\times\mathbb{R}^{k_2}$ is a $k$-vector of Lagrange multipliers (see Rockafellar 1970, chapter 28). 
Assumptions \ref{ass5.1}-\ref{ass5.6} remain unchanged except for Assumption \ref{ass5.3} \textit{(iii)}, which is replaced by:

    \setcounter{assum}{2}
    \begin{assum}   
      \textit{(iii)} There is some $\delta >0$ such that for all $\phi\in B(\phi_{0},\delta)$ there exists a $\theta\in\Theta$ such that $\Psi_{i}(\theta,\phi)<0$, $\forall i=1,\ldots, k_1$.
    \end{assum}

The results of Section \ref{ss_5.2} are still valid with minor modifications in the proofs. We detail these modifications in Appendix \ref{ss_E.4}.

\section{Bayesian Credible Sets }\label{sbcs}
Inferences can be carried out through finite-sample Bayesian credible sets (BCS's).   We study  two kinds of BCS's: credible sets for $\theta$ and credible sets for the identified set $\Theta(\phi)$. 

\subsection{Credible set for $\Theta(\phi)$}

\subsubsection{Two-sided  BCS}
We focus on the case when the identified set is convex and closed, and aim at constructing two-sided credible sets $A_1$ and $A_2$ such that
$$
P(A_1\subset\Theta(\phi)\subset A_2|D_n)\geq 1-\tau
$$
 for $\tau\in (0,1)$, where the probability is taken with respect to the posterior of $\phi$. 
Our construction is based on the support function. To illustrate why support function can help, for a set $\Theta(\phi)$ recall its $\epsilon$-envelope: $\Theta(\phi)^{\epsilon}=\{\theta\in\Theta: d(\theta, \Theta(\phi))\leq\epsilon\}$ and its $\epsilon$-contraction: $\Theta(\phi)^{-\epsilon}=\{\theta\in\Theta(\phi): d(\theta, \Theta\backslash\Theta(\phi))\geq\epsilon\}$ where $\epsilon\geq 0$.  
Let $\hat{\phi}$ be a Bayesian estimator for $\phi_0$,   which can be, e.g., the posterior mean or mode.      We have, for any $c_n\geq0$,
$$
P(\Theta(\hat\phi )^{-c_n}\subset\Theta(\phi)\subset\Theta(\hat\phi )^{c_n}|D_n)=P(\sup_{\|\nu\|=1}|S_{\phi}(\nu)-S_{\hat\phi}(\nu)|\leq c_n|D_n).
$$
Note that the right hand side of the above equation depends on the posterior of the support function.  Let $q_{\tau}$ be the $1-\tau$ quantile of the posterior of $$J(\phi)=\sqrt{n}\sup_{\|\nu\|=1}|S_{\phi}(\nu)-S_{\hat\phi }(\nu)|$$ so that
\begin{equation}\label{eq6.2}
P\left(J(\phi)\leq q_{\tau}\bigg|D_n\right)=1-\tau.
\end{equation}
The posterior of $J(\phi)$ is determined by that of $\phi$. Hence $q_{\tau}$ can be simulated efficiently from the MCMC draws from $p(\phi|D_n)$. Immediately, we have the following theorem:

\begin{thm}\label{th6.2}Suppose  for any $\tau\in(0,1),$  $q_{\tau}$ is defined as  in (\ref{eq6.2}),  then for every sampling sequence $D_n,$
$$
P(\Theta(\hat\phi )^{-q_{\tau}/\sqrt{n}}\subset\Theta(\phi)\subset\Theta(\hat\phi )^{q_{\tau}/\sqrt{n}}|D_n)= 1-\tau.
$$
In particular, $\Theta(\hat\phi )^{-q_{\tau}/\sqrt{n}}$ is allowed to be an empty set.
\end{thm}

\begin{remark}\label{rem_6.3}
It is straightforward to construct the one-sided BCS for $\Theta(\phi)$ using the described procedure. For example, let $\tilde{q}_{\tau}$   be such that $$P( \sqrt{n}\sup_{\|\nu\|=1}(S_{\phi}(\nu)-S_{\hat\phi }(\nu))\leq \tilde{q}_{\tau}|D_n) = 1 - \tau.$$  Then, $P(\Theta(\phi)\subset \Theta(\hat\phi )^{\tilde{q}_\tau/\sqrt{n}}|D_n)=1-\tau$ for every sampling sequence $D_n.$
\end{remark}

\subsubsection{Frequentist coverage probability of BCS for $\Theta(\phi)$}

The constructed two-sided BCS for the identified set has desired frequentist properties, which follows from the Bernstein-von Mises Theorem   (see Theorem \ref{th5.3}) of the support function.
The frequentist coverage probability for a  general (two-sided) multi-dimensional BCS has been largely unknown in the literature before.
The analysis relies on the following assumption, which requires the asymptotic normality and semi-parametric efficiency of the consistent estimator $\hat \phi$. Under  mild conditions, it  holds for many regular estimators such as the posterior mean, mode and the maximum likelihood estimator.

\begin{assum}\label{ass6.2}
The consistent estimator $\hat\phi $ satisfies $$
\sqrt{n}(\hat\phi -\phi_0)\rightarrow^d \mathcal{N}(0, I_0^{-1})
$$
where $I_0$ denotes the semi-parametric efficient information matrix as in Assumption \ref{ass5.9}.
\end{assum}

\begin{thm} \label{th6.3} Consider the moment inequality model in (\ref{eq4.1})-(\ref{eq4.2old}). If the Assumptions of Theorem \ref{th5.3} and Assumption \ref{ass6.2} hold,  then the constructed two-sided Bayesian credible set has asymptotically correct frequentist coverage probability,  that is, for any $\tau\in(0,1)$,
$$
P_{D_n}(\Theta(\hat\phi )^{-q_{\tau}/\sqrt{n}}\subset\Theta(\phi_0)\subset\Theta(\hat\phi )^{q_{\tau}/\sqrt{n}})\geq 1-\tau+o_p(1).\footnote{The result presented here is understood as: There is a random sequence $\Delta(D_n)$ that depends on $D_n$  such that $\Delta(D_n)=o_p(1)$, and for any sampling sequence $D_n$, we have $P_{D_n}(\Theta(\hat\phi )^{-q_{\tau}/\sqrt{n}}\subset\Theta(\phi_0)\subset\Theta(\hat\phi )^{q_{\tau}/\sqrt{n}})\geq 1-\tau+\Delta(D_n)$. Similar interpretation applies to (\ref{eq6.3add}).}
$$
where $P_{D_n}(.)$ denote the probability measure based on the sampling distribution, fixing $(\theta, \phi)=(\theta_0,\phi_0)$.
\end{thm}

Note that in   Theorem \ref{th6.2},   the random set is $\Theta(\phi)$, while in Theorem \ref{th6.3} the random sets are  $\Theta(\hat\phi )^{-q_{\tau}/\sqrt{n}}$ and $\Theta(\hat\phi )^{q_{\tau}/\sqrt{n}}$. 
 The rationale of this theorem is that, because the identified set itself is ``point identified", its prior does not depend on that of $\theta$ and is   dominated by the data asymptotically.
  
\begin{remark}
Note that $q_{\tau}$ depends only on the posterior of $\phi$.  Hence Theorem \ref{th6.3} does not rely on the  prior of $\theta$, and shows asymptotic robustness to the prior of $\phi$. It    also holds when $\Theta(\phi)$ becomes a singleton, and in that case the lower-side $\Theta(\hat\phi )^{-q_{\tau}/\sqrt{n}}$ is empty. Therefore the point identified case is also nested. We shall discuss the uniformity issue in more detail in Section \ref{s_further_illustration_Uniformity}.
\end{remark}

Similarly, we can show that the one-sided BCS as constructed in Remark \ref{rem_6.3} above has asymptotically correct coverage probability too.  For example, for $\tilde{q}_{\tau}$  such that \\ $P(\sqrt{n}\sup_{\|\nu\|=1}(S_{\phi}(\nu)-S_{\hat\phi }(\nu))\leq \tilde{q}_{\tau}|D_n)=1-\tau$,  then
\begin{equation}\label{eq6.3add}
P_{D_n}(\Theta(\phi_0)\subset\Theta(\hat\phi )^{\tilde{q}_{\tau}/\sqrt{n}})\geq 1-\tau+o_p(1).
\end{equation}



\subsection{Credible set for $\theta$}
We now construct the Bayesian credible set for $\theta$. A BCS for $\theta$ at level $1-\tau$ is a set BCS$(\tau)$ such that
\begin{equation*}\label{eq6.1}
P(\theta\in \bcs(\tau)|D_n)=1-\tau
\end{equation*}
for $\tau\in(0,1)$.  One of the popular choices of the credible set is the highest-probability-density (HPD) set, which has been widely used in empirical studies and also used in the Bayesian partially identified literature by e.g., Moon and Schorfheide (2012) and Norets and Tang (2012). 

The BCS  can be compared with the frequentist confidence set (FCS).  A frequentist confidence set FCS$(\tau)$ for $\theta_0$  satisfies
$$
\lim_{n\rightarrow\infty}\inf_{\phi\in\Phi}\inf_{\theta_{0}\in\Theta(\phi)}P_{D_n}(\theta_{0}\in \fcs(\tau))\geq 1-\tau.
$$
There have been various procedures in the literature to construct a FCS$(\tau)$ that satisfies the above inequality. One of the important  FCS's  is   based on a consistent estimator $\hat{\phi}$ of $\phi_0$ such that $\Theta(\hat{\phi})\subset\fcs({\tau})$. By using a known likelihood function, Moon and Schorfheide (2012) compared the BCS with this type of FCS and showed that the BCS and FCS are asymptotically different. As Theorem \ref{t6.1} below shows, such a comparison still carries over under the more robust semi-parametric Bayesian setup.  The following assumption is needed.
\begin{assum}\label{ass6.1}
(i) The frequentist FCS($\tau$) is such that, there is $\hat{\phi}$ with $\|\hat{\phi}-\phi_0\|=o_p(1)$ satisfying $\Theta(\hat{\phi})\subset \fcs(\tau)$.\\
(ii) $\pi(\theta\in\Theta(\phi)|\phi)=1$ for all $\phi\in\Phi$; $\sup_{(\theta,\phi)\in\Theta\times\Phi}\pi(\theta|\phi)<\infty$.
\end{assum}
 Many frequentist FCS's satisfy condition \textit{(i)}, see, e.g., Imbens and Manski (2004), Chernozhukov et al. (2007), Rosen (2008), Andrews and Soares (2010), etc. Condition \textit{(ii)} is easy to verify since $\Theta\times\Phi$ is compact. When for every $\phi\in\Phi$, $\Theta(\phi)$ is not a singleton, examples of $\pi(\theta|\phi)$ satisfying assumption \ref{ass6.1} \textit{(ii)} include: the uniform prior with density $$\pi(\theta|\phi)=\mu(\Theta(\phi))^{-1}I_{\theta\in\Theta(\phi)},$$ where $\mu(\cdot)$ denotes the Lebesgue measure; and the truncated normal prior with density
$$\pi(\theta|\phi)=\left[\int_{\Theta(\phi)}h(x;\lambda,\Sigma)dx\right]^{-1}h(\theta; \lambda,\Sigma)I_{\theta\in\Theta(\phi)},$$
where $h(x;\lambda,\Sigma)$ is the density function of a multinormal distribution $\mathcal{N}(\lambda,\Sigma)$.

\begin{thm}\label{t6.1}
Under Assumption \ref{ass6.1} and  the assumptions of Theorem \ref{th5.2},   $\forall\tau\in(0,1)$,\\
(i)
$$
P(\theta\in\fcs(\tau)|D_n)\rightarrow^p1,
$$
(ii) $$
P(\theta\in\fcs(\tau), \theta\notin\bcs(\tau)|D_n)\rightarrow^p\tau.
$$
\end{thm}
\begin{remark}

Theorem \ref{t6.1} \textit{(i)} shows that the posterior probability that $\theta$ lies inside the frequentist confidence set is arbitrarily close to one, as $n\rightarrow\infty$. This indicates that the  posterior will asymptotically concentrate inside the FCS. On the other hand, by  \textit{(ii)},  there is  a non-negligible probability that  FCS is strictly larger than BCS.  The prior information on $\theta$ still plays a non-negligible role in the posterior as the sample size increases.

Our prior condition in Assumption \ref{ass6.1} \textit{(ii)} implies that Theorem \ref{t6.1} only focuses on partial identification. It can be restrictive in the point identified case. Because our prior is such that $\pi(\theta\in\Theta(\phi)|\phi)=1$ for each $\phi$, when $\Theta(\phi)$ is a singleton $\pi(\theta|\phi)$ becomes a Dirac function and $\sup_{\theta,\phi}\pi(\theta|\phi)<\infty$  cannot be expected to hold in this case. On the other hand, Assumption \ref{ass6.1} \textit{(ii)} does cover many partially identified models of interest, and it is a prior assumption that has been used frequently elsewhere in the literature, e.g., Moon and Schorfheide (2012) and Gustafson (2012).
\end{remark}

\section{Projection and Subset Inference}\label{s_Projection_Subset_Inference}
One of the important features of the proposed   procedure is that it is relatively easy to marginalize onto low-dimensional subspaces, and the computation is fast. Suppose the dimension of $\theta$ is relatively large, but we are interested in only one component of $\theta$, say $\theta_1$. Then projections aim at constructing the BCS's for $\theta_1$ and for its identified set  $\widetilde{\Theta}(\phi)_1$.

We illustrate this by using the interval regression example (Example \ref{ex2.2}). Suppose the full parameter $\theta$ is high-dimensional. Let $W_1=ZY_1$, $W_2=ZY_2$ and $V=Zx^T$. Here $\phi_1=EW_1$,  $\phi_2=EV$ and $\phi_3=EW_2$. Let $\phi=(\phi_1^T,vec(\phi_2)^T,\phi_3^T)$, and $e=(1,0,...,0)^T$. The identified set for $\theta_1$ can be expressed using the support function $S_{\phi}(\cdot)$:
$$
\widetilde{\Theta}(\phi)_1=\{\theta_1: \exists  \omega=(\theta_2,...,\theta_{d})\text{ such that } (\theta_1,\omega)\in\Theta(\phi)\}=[-S_{\phi}(-e), S_{\phi}(e)],
$$
where the exact expression for $S_{\phi}(\cdot)$ is given in Appendix C.1 in the supplementary material. We place a Dirichlet process prior $ \mathcal{D}ir(\nu_{0},Q_{0})$ on the joint CDF of $(W_1, W_2, V)$. By the stick-breaking representation (see Sethuraman 1994) the deduced   posterior distribution of $\phi$ is the distribution of the following quantity:
\begin{equation}\label{e6.4}
       \phi|D_{n}= \rho\sum_{i=1}^{n}\beta_{i}D_{n,i} + (1 - \rho)\sum_{j=1}^{\infty}\alpha_{j}\xi_{j}
\end{equation}
 where $D_{n,i}$ is the $i$th observation of the vector $(W_1^T, vec(V)^T, W_2^T)$, $\rho$ is drawn from a Beta distribution $\mathcal{B}e(n, \nu_{0})$ independently of the other quantities, $(\beta_{1},\ldots,\beta_{n})$ is drawn from a Dirichlet distribution of parameters $(1,\ldots,1)$ on the simplex $S_{n-1}$ of dimension $(n-1)$, $\xi_{j}\sim \:iid\: Q_{0}$ and $\{\alpha_{k}\}_{k\geq 1}$ are computed as $\alpha_{k} = v_{k}\prod_{l=1}^{k}(1 - v_{l})$ where $\{v_{l}\}_{l\geq 1}$ are independent drawings from a beta distribution $\mathcal{B}e(1,\nu_{0})$ and $\{v_{j}\}_{j\geq1}$ are independent of $\{\xi_{j}\}_{j\geq 1}$. In practice, we can set a  truncation $K$, so the infinite sum in the posterior representation in (\ref{e6.4}) is replaced with a truncated sum $(1-\rho)\sum_{j=1}^K\alpha_j\xi_j.$ In addition, $(\alpha_1,...\alpha_K)$ are normalized so that $\sum_{j=1}^K\alpha_j=1.$

 We can place a uniform prior for $\theta$, and draw $\{\theta^{(i)},\phi^{(i)}\}_{i=1}^B$ from the posterior  $(\theta, \phi)|D_n$. Then $\{\theta_1^{(i)}\}_{i=1}^B$ are the draws from the marginal  posterior of $\theta_1.$  Let $\theta_1^{(\tau/2)}$ and $\theta_1^{(1-\tau/2)}$ be the $\tau/2$th and  ${(1-\tau/2)}$th sample quantiles of $\{\theta_1^{(i)}\}_{i=1}^B$. Then $[\theta_1^{(\tau/2)}, \theta_1^{(1-\tau/2)}]$ is the BCS($\tau$) of $\theta_1$. Moreover, let $q_{\tau}$ be the $(1-\tau)$th quantile of the posterior of
$$
J(\phi)=\sqrt{n}\max\{S_{\phi}(e)-S_{\hat\phi }(e), S_{\phi}(-e)-S_{\hat\phi }(-e)\},
$$
which can be approximated by the $(1-\tau)$th sample quantile of $\{J(\phi^{(i)})\}_{i=1}^B$.
We then construct the BCS$(\tau)$ for $\widetilde{\Theta}(\phi)_1$ as
$
[-S_{\hat\phi }(-e)-\frac{q_{\tau}}{\sqrt{n}},  S_{\hat\phi }(e)+\frac{q_{\tau}}{\sqrt{n}}].
$


We present a simple numerical result for illustration, where $\theta_{01}=1$, but the total dimension is high: $\dim(\theta_0)=10.$ Let $W_1\sim \mathcal{N}(0,0.5I)$ and $W_2\sim \mathcal{N}(5, I)$.  Set $\nu_0=3$ and the base measure $Q_0=\mathcal{N}(0, I)$.  $B=100$ posterior draws are sampled.  While the finite sample performance is very robust to the choice of the truncation $K$,  we choose $K$ following the guidance of Ishwaran and James (2002), who obtained an approximation error of order $n\exp(-(K-1)/\nu_0)$ for truncations. Hence, in the simulation with $n=500, \nu_0=3$, the choice $K=50$ gives  an error of order $4\times 10^{-5}$. Table \ref{table1} summarizes the true identified set $\widetilde\Theta(\phi_0)_1$ for $\theta_1$, and the averaged BCS$(0.1)$ for both $\theta_1$ and the projected set $\widetilde{\Theta}(\phi)_1$ over 50 replications. Results based on various choices of $(n, B, K)$ are reported.

\begin{table}[htdp]
\caption{90\% Bayesian credible sets  marginalized to the subset for $\theta_1$}
\begin{center}

\begin{tabular}{cccccc}
\hline
$n$ & $K$ & $B$ &   $\widetilde\Theta(\phi_0)_1$& BCS for $\theta_1$ & BCS for $\widetilde{\Theta}(\phi)_1$ \\
\hline
500 & 50 & 100 & [0,    1.667] & [0.007,    1.667] & [-0.174,    1.844] \\
 & 50 & 500 &  & [-0.008    1.666] & [-0.174,    1.837] \\
 & 100 & 100 &  & [-0.012    1.662] & [-0.181,    1.832] \\
 & 100 & 500 &  & [-0.000    1.667] & [-0.169,    1.840] \\
 \hline
1000 & 50 & 100 &  [0,    1.667] & [0.023,   1.641] & [-0.121,    1.789] \\
 & 50 & 500 &  & [0.011,    1.641] & [-0.126,  1.786] \\
 & 100 & 100 &  & [0.036,    1.636] & [-0.120,    1.781] \\
 & 100 & 500 &  & [0.025,   1.641] & [-0.121,    1.786] \\
\hline
\end{tabular}

\label{table1}
  \end{center}
\end{table}

When computing the BCS for $\widetilde{\Theta}(\phi)_1$,  it is also interesting to compare the computation time      with that of the high-dimensional projection based on the criterion function approach as in, e.g. Chernozhukov et al. (2007), Andrews and Soares (2010) etc, because they have the same asymptotic frequentist coverages as ours.  For the moment inequalities $\Psi(\theta,\phi)=(\phi_2\theta-\phi_3,\phi_1-\phi_2\theta)\leq0$ we employ the criterion function and construct a confidence set FCS as in Chernozhukov et al. (2007):  $$Q_n(\theta)=\sum_{j}\max(\Psi_j(\theta,\hat\phi),0)^2w_j,\quad \text{FCS}(\tau)=\{\theta: \sqrt{n}Q_n(\theta)\leq c_{\tau}\},
$$  with $w_j=1$ and $\hat\phi$ the sample mean estimator of $\phi$. The critical value $c_{\tau}$ is obtained via the bootstrap procedure proposed by Bugni (2010), which  requires solving a constrained optimization problem. We  use     the ``fmincon" toolbox in Matlab for the numerical optimization\footnote{We use the Matlab code of Bugni (2010), downloaded from the online supplement of Econometrica. The optimization is solved constrained on an estimated identified set, which involves an additional parameter $t_n$. We set $t_n=\log(n).$}, and then project  FCS$(\tau)$ onto the subspace for $\theta_1$ to get the marginal confidence set FCS$_1(\tau)$. The projection is done through the following steps: generate $\{\theta_j^*\}_{j=1}^M$ uniformly from $\Theta$. Let $\theta_{j,1}^*$ be the first component of $\theta_j^*$ and
\begin{equation}\label{eq7.2}
L(\tau)=\min\{\theta_{j,1}^*:  \theta_{j}^*\in\text{FCS}(\tau), j=1,...,M\},\quad U(\tau)=\max\{\theta_{j,1}^*:  \theta_{j}^*\in\text{FCS}(\tau), j=1,...,M\}.
\end{equation}
Then $[L(\tau), U(\tau)]$ forms a projected frequentist confidence interval for $\widetilde\Theta(\phi_0)_1.$  In the simulation, we set a small parameter space $\Theta=\otimes_{i=1}^{10}[-2, 2]$ in order to calculate $L(\tau)$ and $U(\tau)$ efficiently.

Table  \ref{table2add} compares the computation times necessary to obtain the projected sets using our proposed BCS and using the criterion function approach. Reported is the averaged time for one computation  over 50 replications, using the same simulated model.  We see that the proposed BCS projection computes much faster.

\begin{table}[htdp]
\caption{Computation times (in seconds) for the projected BCS and criterion-function-FCS }
\begin{center}

\begin{tabular}{cc|ccc|ccc}
\hline
 &  &  & BCS   &  & & FCS& \\
 &  &  & $K$        &  & &$M$ \\
$n$ & $B$ & 50 & 100 & 500 & 30 & 50 & 100 \\
\hline
 &  &  &         &  & & \\
500 & 50 & 0.065 & 0.073 & 0.129 & 9.169 & 10.214 & 10.955 \\
 & 100 & 0.128 & 0.142 & 0.248 & 18.963 & 18.893 & 19.496 \\
 & 200 & 0.244 & 0.273 & 0.479 & 37.067 & 36.599 & 37.288 \\
  &  &  &         &  & & \\
1000 & 50 & 0.072 & 0.079 & 0.136 & 14.123 & 14.248 & 15.837 \\
 & 100 & 0.137 & 0.155 & 0.259 & 26 & 27.029 & 28.442 \\
 & 200 & 0.269 & 0.295 & 0.549 & 54.027 & 52.661 & 54.240 \\
\hline
\end{tabular}

\label{table2add}
  \end{center}
 \small \textit{Proposed BCS and criterion-function-based-FCS are compared. $B$ is the number of either posterior draws (for BCS) or Bootstrap draws (for FCS)   to  compute the critical values; $K$ is the truncation number to approximate the Dirichlet process posterior; $M$ is used in (\ref{eq7.2}) for  the projected FCS. Computations are conducted using a 2.3 GHz Mac with Intel Core i7 CPU.}

\end{table}


\section{Posterior consistency for $\Theta(\phi)$}\label{s_posterior_consistency_set}

 The estimation accuracy of the identified set is often measured, in the literature, by the Hausdorff distance. Specifically, for a point $ a$ and a set $A$, let
$
d(a,A)=\inf_{x\in A}\|a-x\|,
$
where $\|\cdot\|$ denotes the Euclidean norm. The Hausdorff distance between sets $A$ and $B$ is defined as
\begin{equation*}
d_H(A, B)= \max\left\{\sup_{a\in A}d(a, B), \sup_{b\in B}d(b, A)\right\}=\max\left\{\sup_{a\in A}\inf_{b\in B}\|a-b\|, \sup_{b\in B}\inf_{a\in A}\|b-a\|\right\}.
\end{equation*}
  This section aims at deriving a rate $r_n=o(1)$ such that for some constant $C>0,$
$$
P(d_H(\Theta(\phi), \Theta(\phi_0))<Cr_n|D_n)\rightarrow^p1.
$$
The above result is based upon two important features: (1) the posterior concentration rate for $\phi$ which is stated in assumption \ref{ass4.1}, and (2) the continuity of the Hausdorff distance with respect to $\phi$. 



The continuity of $d_{H}(\Theta(\phi),\Theta(\phi_{0}))$ with respect to $\phi$ in multi-dimensional models is hard to verify. Hence, instead of assuming the continuity directly, we place a less demanding assumption which implicitly implies the continuity but is relatively easier to verify. With this aim, we consider the \textit{moment inequality model} described in equations (\ref{eq4.1}) - (\ref{eq4.2old}) and place the following assumptions.

\begin{assum}\label{ass8.1}
The parameter space $\Theta\times\Phi$ is compact.
\end{assum}

\begin{assum} \label{ass4.3}
$\{\Psi(\theta,\cdot): \theta\in\Theta\}$ is Lipschitz equi-continuous on $\Phi$, that is, for some $K>0$, $\forall \phi_1, \phi_2\in \Phi$,
$$
\sup_{\theta\in\Theta}\|\Psi(\theta, \phi_1)-\Psi(\theta, \phi_2)\|\leq K\|\phi_1-\phi_2\|.
$$
\end{assum}
Given the compactness of $\Theta$, this assumption is satisfied by many interesting examples of  moment inequality models.

\begin{assum} \label{ass4.4}
There exists a closed neighborhood $U(\phi_0)$ of $\phi_0$, such that for any $a_n=O(1)$, and any $\phi\in U(\phi_0)$, there exists $C>0$ that might depend on $\phi$, so that
$$
\inf_{\theta: d(\theta, \Theta(\phi))\geq Ca_n}\max_{i\leq k} \Psi_i(\theta,\phi)>a_n.
$$
\end{assum}

Intuitively, when $\theta$ is bounded away from $\Theta(\phi)$ (up to a rate $a_n$), at least one of the moment inequalities is violated, which means $\max_{i\leq k}\Psi_i(\theta,\phi)>0$. This assumption quantifies how much $\max_{i\leq k}\Psi_i(\theta,\phi)$ will depart from zero. This is a sufficient condition for the partial identification condition (4.5) in Chernozhukov, Hong and Tamer (2007). If we define $$
Q(\theta,\phi)=\|\max(\Psi(\theta,\phi),0)\|= \left[\sum_{i=1}^k(\max(\Psi_i(\theta,\phi), 0))^2\right]^{1/2}
$$
then $Q(\theta,\phi)=0$ if and only if $\theta\in\Theta(\phi)$. The partial identification condition in Chernozhukov et al. (2007, condition (4.5)) assumes that  there exists $K>0$ so that for all $\theta$,
\begin{equation}\label{eq4.2}
Q(\theta,\phi)\geq  Kd(\theta,\Theta(\phi)),
\end{equation}
which says that $Q$ should be bounded below by a number proportional to the distance of $\theta$ from the
identified set if $\theta$ is bounded away from the identified set.  Assumption \ref{ass4.4} is a sufficient condition for (\ref{eq4.2}).
\begin{exm}[Interval censored data - \textit{continued}] In the interval censoring data example, $\Psi(\theta, \phi)=(\theta-\phi_2, \phi_1-\theta)^T$ and for any $\phi=(\phi_1, \phi_2)$ and $\tilde{\phi}=(\tilde{\phi}_1, \tilde{\phi}_2)$ we have:
$\|\Psi(\theta, \phi)-\Psi(\theta, \tilde{\phi})\|=\|\phi-\tilde{\phi}\|$. This verifies Assumption \ref{ass4.3}. Moreover,   for any $\theta$ such that $d(\theta,\Theta(\phi))\geq a_n$, either $\theta\leq\phi_1-a_n$ or $\theta\geq\phi_2+a_n.$ If $\theta\leq \phi_1-a_n$, then $\Psi_2(\theta,\phi)=\phi_1-\theta\geq a_n$; if $\theta\geq \phi_2+a_n$, then $\Psi_1(\theta,\phi)=\theta-\phi_2\geq a_n.$ This verifies Assumption \ref{ass4.4}. $\square$
\end{exm}

The following theorem shows the concentration rate for the posterior of the identified set.
\begin{thm}\label{t4.1}
Under Assumptions \ref{ass4.1}, \ref{ass8.1}-\ref{ass4.4}, for some $C>0$,
\begin{equation}\label{eq_t4.1}
  P(d_H(\Theta(\phi), \Theta(\phi_0))>C\sqrt{\frac{\log n}{n}}|D_n)\rightarrow^p0.
\end{equation}
\end{thm}
\begin{remark}
The convergence in Hausdorff distance can be implied by that of the support function for convex and close sets (e.g., Beresteanu and Molinari 2008).  Therefore, (\ref{eq_t4.1}) is another statement of result (\ref{eq5.2}). However, they are obtained under different assumptions   and  (\ref{eq_t4.1}) is obtained directly from the perspective of the posterior of the identified set. 
\end{remark}
\begin{remark}
Recently, Kitagawa (2012) obtained the posterior consistency for $\Theta(\phi)$ in the one-dimensional case:
$
P(d_H(\Theta(\phi), \Theta(\phi_0))>\epsilon|D_n)\rightarrow0
$
for almost every sampling sequence of $D_n.$ This result was obtained for the case where  $\Theta(\phi)$ is a connected interval and $d_H(\Theta(\phi), \Theta(\phi_0))$ is assumed to be a continuous map of $\phi$. In multi-dimensional cases where $\Theta(\phi)$ is a more general convex set, however, verifying the continuity of  $d_H(\Theta(\phi), \Theta(\phi_0))$  is much more technically involved, due to the challenge of computing the Hausdorff distance in multi-dimensional manifolds. In contrast, our Lipschitz equi-continuity condition in Assumption \ref{ass4.3} and Assumption \ref{ass4.4} are  much easier to verify in specific examples, as they depend on the moment conditions directly.
\end{remark}

\section{Further Illustrations and  Uniformity}\label{s_further_illustration_Uniformity}
\subsection{Missing data: coverage probabilities and prior sensitivity}\label{smd}
This subsection illustrates the coverages of the proposed BCS in the missing data problem (example \ref{ex2.3}),  previously discussed  by Manski (2003).   Let $Y$ be a binary variable, indicating whether a treatment is successful ($Y=1$) or not ($Y=0$).  However, $Y$ is observed subject to missing. We write $M=0$ if $Y$ is missing, and $M=1$ otherwise. Hence, we observe $(M, MY)$. The parameter of interest is $\theta = P(Y=1)$. The identified parameters are denoted by
$$
\phi_1=P(M=1),\quad \phi_2=P(Y=1|M=1).
$$ Let $\phi_0=(\phi_{10},\phi_{20})$ be the true value of $\phi=(\phi_1,\phi_2)$.
Then, without further assumption on $P(Y=1|M=0)$, $\theta$ is only partially identified on $
\Theta(\phi)=[\phi_1\phi_2,\phi_1\phi_2+1-\phi_1]
$.
The support function is easy to calculate and is
$$S_{\phi}(1)=\phi_1\phi_2+1-\phi_1\quad S_{\phi}(-1)=-\phi_1\phi_2.$$ Suppose we observe i.i.d. data  $\{(M_i, Y_iM_i)\}_{i= 1}^{n}$, and define $\sum_{i=1}^nM_i=n_1$ and $\sum_{i=1}^nY_iM_i=n_2$. In this example, the true likelihood function
$
l_n(\phi) \propto \phi_1^{n_1}(1-\phi_1)^{n-n_1}\phi_2^{n_2}(1-\phi_2)^{n_1-n_2}
$ is known.

We place independent beta priors, Beta($\alpha_1,\beta_1$) and Beta$(\alpha_2,\beta_2)$, on $(\phi_1,\phi_2)$. The uniform distribution is a special case of Beta prior. Then the posterior of $(\phi_1,\phi_2)$ is a product of Beta($\alpha_1+n_1,\beta_1+n-n_1$) and Beta($\alpha_2+n_2,\beta_2+n_1-n_2$). If in addition, we have  prior information on $\theta$ and place a prior $\pi(\theta|\phi)$ supported on $\Theta(\phi)$, then by integrating out $\phi$, we immediately obtain the marginal posterior of $\theta.$

We now present the two-sided BCS for $\Theta(\phi)$ obtained by using the support function of $\Theta(\phi)$. The estimator $\hat\phi$ is taken as the posterior mode: $
\hat{\phi}_{1}= (n_1+\alpha_1-1)/(n+\alpha_1+\beta_1-2),$ and $\hat{\phi}_{2}=(n_2+\alpha_2-1)/(n_1+\alpha_2+\beta_2-2)$.
Then
$$J(\phi)=\sqrt{n}\max\left\{|\phi_1\phi_2-\phi_1-\hat{\phi}_{1}\hat{\phi}_{2}+\hat{\phi}_{1}|, |\phi_1\phi_2-\hat{\phi}_{1}\hat{\phi}_{2}|\right\}.
$$
Let $q_{\tau}$ be  the $1-\tau$ quantile of the posterior of $J(\phi)$, which can be obtained by simulating from the Beta distributions. The lower and upper $1-\tau$ level BCS's  for $\Theta(\phi)$ are $\Theta(\hat\phi )^{-q_{\tau}/\sqrt{n}}\subset \Theta(\phi)\subset \Theta(\hat\phi )^{q_{\tau}/\sqrt{n}}$ where
$$
\Theta(\hat\phi )^{-q_{\tau}/\sqrt{n}}=[\hat{\phi}_{1}\hat{\phi}_{2}+q_{\tau}/\sqrt{n}, \hat{\phi}_{1}\hat{\phi}_{2}+1-\hat{\phi}_{1}-q_{\tau}/\sqrt{n}]
$$
\noindent and
$$
\Theta(\hat\phi )^{q_{\tau}/\sqrt{n}}=[\hat{\phi}_{1}\hat{\phi}_{2}-q_{\tau}/\sqrt{n}, \hat{\phi}_{1}\hat{\phi}_{2}+1-\hat{\phi}_{1}+q_{\tau}/\sqrt{n}],
$$
which are also two-sided asymptotic $1-\tau$ frequentist confidence intervals of the true $\Theta(\phi_{0}).$

Here we present a simple simulated example, where the true $\phi_0=(0.7,0.5)$. This implies the true identified interval to be $[0.35, 0.65]$ and about thirty percent of the simulated data are ``missing". We   set  $\alpha_1=\alpha_2, \beta_1=\beta_2$ in the prior. In addition, $B=1,000$ posterior draws $\{\phi^i\}_{i=1}^B$ are sampled from  the posterior Beta distribution. For each of them, compute $J(\phi^{i})$ and set $q_{0.05}$ as the 95\% upper quantile of   $\{J(\phi^i)\}_{i=1}^B$ to obtain the critical value of the BCS and construct the two-sided BCS for the identified set. Each simulation is repeated for 500 times to calculate the coverage frequency of the true identified interval.  Table \ref{table2} presents the results. We see that the coverage probability for the two-sided is close to the desired 95\% when sample size increases. In addition, the marginal coverages of the lower and upper sets are close to 97.5\% when sample size is relatively large.

Moreover, Figure 1 plots the five conjugate prior  specifications  used in this study: flat prior, reverse $J$-shaped with a right tail, $J$-shaped with a left tail,  $U$-shaped, and uni-mode. These priors reflect different types of prior beliefs: the first prior is used if a researcher has no informative prior information, the second one (resp. third one) is used if one strongly believes that the probability of missing is low (resp. high), the fourth prior is used when one   thinks  that the probability of missing is either very high or very low, and the last prior corresponds to a  symmetric prior belief centered at fifty percent. So Table \ref{table2} also provides   simple sensitivity analysis of prior specification. The results demonstrate robustness of the coverage probabilities to conjugate prior specification.

\begin{table}[htdp]
\caption{Frequentist coverage probability of  BCS and prior sensitivity for missing data}
\begin{center}

\begin{tabular}{c|cc|ccc}
\hline
$n$ & $\alpha$ & $\beta$ & Lower & Upper & Two-sided \\
\hline
50 & 1 & 1 & 0.978 & 0.944 & 0.924 \\
 & 1 & 0.1 & 0.964 & 0.944 & 0.912 \\
 & 0.1 & 1 & 0.952 & 0.958 & 0.916 \\
 & 0.1 & 0.1 & 0.974 & 0.958 & 0.938 \\
   &2&2&0.958 &  0.970& 0.932\\
 \hline
100 & 1 & 1 & 0.982 & 0.96 & 0.948 \\
 & 1 & 0.1 & 0.978 & 0.968 & 0.950 \\
 & 0.1 & 1 & 0.968 & 0.968 & 0.948 \\
 &0.1  & 0.1 & 0.972 & 0.972 & 0.944 \\
  &2&2&0.956 &  0.978& 0.944\\
 \hline
500 & 1 & 1 & 0.970 & 0.974 & 0.950 \\
 &  1& 0.1 & 0.978 & 0.978 & 0.958 \\
 & 0.1 & 1 & 0.974 & 0.972 & 0.948 \\
 & 0.1 & 0.1 & 0.972 & 0.974 & 0.950 \\
 &2&2& 0.976 & 0.978& 0.956\\
\hline
\end{tabular}

\label{table2}
\small

\it Lower, Upper and Two-sided represent the  frequencies of the events $\Theta(\hat\phi )^{-q_{\tau}/\sqrt{n}}\subset\Theta(\phi_0)$, $\Theta(\phi_0)\subset \Theta(\hat\phi )^{q_{\tau}/\sqrt{n}}$, and $\Theta(\hat\phi )^{-q_{\tau}/\sqrt{n}}\subset\Theta(\phi_0)\subset \Theta(\hat\phi )^{q_{\tau}/\sqrt{n}}$ over 500 replicates. The coverage probability for the two-sided BCS is set to 95\%.

\end{center}
\end{table}

 \begin{figure}[htbp]
\begin{center}
\caption{Conjugate priors   for sensitivity analysis in the missing data problem}
\includegraphics[width=12cm]{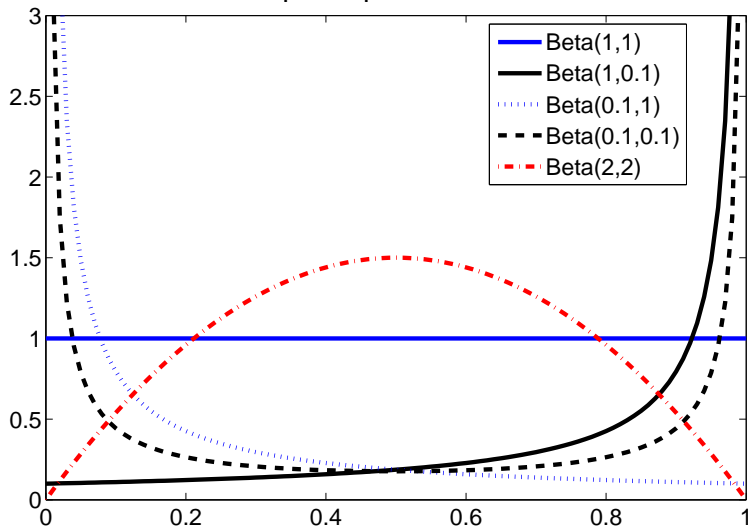}
\label{figureadd}
\end{center}
\end{figure}


\subsection{Uniformity: from partial identification to point identification}\label{s_from_partial_identification_point_Identification}

We have been focusing on partially identified models, and the inference results achieved are for a fixed data generating process. It is interesting to see whether they  still hold uniformly over a class of data generating processes, including the case   when point identification is nearly achieved. This is important because in many cases it is possible that we actually have point identification and, in that event, $\Theta(\phi)$ degenerates to a singleton. For example, in the interval censored model, when  $EY_1=EY_2$,  $\theta=EY$ is point identified.

When point identification is indeed achieved, the frequentist coverage probability of the upper-sided BCS $\Theta(\phi)\subset\Theta(\hat\phi )^{q_{\tau}/\sqrt{n}}$ and the asymptotic normality for the posterior of the support function    still hold   because they are  generally guaranteed by the semi-parametric Bernstein-von Mises theorem for $\phi$ when $\Theta(\phi)$ is a singleton (e.g., Rivoirard and Rousseau 2012, Bickel and Kleijn 2011). But the low-side BCS for $\Theta(\phi)$ will be empty with a  positive probability. 
Theorem \ref{t6.1}, however, does not hold anymore  when $\theta$ is point identified, as we discussed previously.
We further illustrate the uniformity in two examples.

\begin{exm}[Interval censored data - \textit{continued}] We  show that the   the upper  BCS for the identified set has a uniformly correct frequentist asymptotic coverage probability.
To simplify our illustration, we assume $Y_1$ and $Y_2$ are independent and follow $\mathcal{N}(\phi_{10}, 1)$ and $\mathcal{N}(\phi_{20},1)$, respectively. A sequence  of different $\phi_0$ are  considered which includes the case $\phi_{10}-\phi_{20}=o(1)$. When $\phi_{10}=\phi_{20}$, however,  we suppose $Y_1, Y_2$ are sampled independently.  Suppose   econometricians   place independent standard normal priors on $\phi_1$ and $\phi_2$,   then the posteriors  are independent, given by $\phi_i|D_n\sim \mathcal{N}(\bar{Y}_i\frac{n}{1+n}, \frac{1}{1+n}), i=1,2,
$
and  $\hat{\phi} = \frac{n}{1+n}(\bar{Y}_1,\bar{Y}_2)$ is the posterior mode of the joint distribution. 
The support function is $S_{\phi}(1)=\phi_2$, $S_{\phi}(-1)=-\phi_1$. Let $$J(\phi)=\sqrt{n}\sup_{\|\nu\|=1}(S_{\phi}(\nu)-S_{\hat\phi }(\nu))=\sqrt{n}\max\left\{\phi_2-\frac{n}{1+n}\bar{Y}_2, \frac{n}{1+n}\bar{Y}_1-\phi_1\right\},$$ and let $\tilde{q}_{\tau}$ be the $1-\tau$ quantile of the posterior of $J(\phi)$.   We now show that the frequentist coverage of  BCS($\tau$)$=[\bar{Y}_1\frac{n}{1+n}-\frac{\tilde{q}_{\tau}}{\sqrt{n}}, \bar{Y}_2\frac{n}{1+n}+\frac{\tilde{q}_{\tau}}{\sqrt{n}}]$ is  valid uniformly for $\phi_0=(\phi_{10},\phi_{20})\in\Phi$, that is,
\begin{equation}\label{e71}
\liminf_{n\rightarrow\infty}\inf_{(\phi_{01},\phi_{02})\in\Phi}P_{D_n}(\Theta(\phi_0)\subset \text{BCS}(\tau))= 1-\tau.
\end{equation}
We can simplify $J(\phi)$ to be $\sqrt{n/(1+n)}\max\{Z_1, Z_2\}$ where $Z_i\sim \mathcal{N}(0,1)$, $i=1,2$ and $Z_1$ and $Z_2$ are independent. This implies, for the standard normal's CDF $H(\cdot)$,
$$
1 - \tau = P(J(\phi)\leq\tilde{q}_{\tau}|D_n)=P\left(\max\{Z_1,Z_2\}\leq\tilde{q}_{\tau}\sqrt{\frac{1+n}{n}}\right)
= H\left(\tilde{q}_{\tau}\sqrt{\frac{1+n}{n}}\right)^2.
$$
Hence, $H\left(\tilde{q}_{\tau}\sqrt{(1+n)/n}\right)^2 = H(\tilde{q}_{\tau})^2 + o(1)$ and so $H(\tilde{q}_{\tau})^2\rightarrow1-\tau.$ The event $\{\Theta(\phi_0)\subset \textrm{BCS}(\tau)\}$ is equivalent to $(\bar{Y}_1-\phi_{10})\frac{n}{1+n}\leq \frac{\tilde{q}_{\tau}}{\sqrt{n}}+\frac{\phi_{10}}{1+n}$ and $(\bar{Y}_2-\phi_{20})\frac{n}{1+n}\geq \frac{\phi_{20}}{1+n}- \frac{\tilde{q}_{\tau}}{\sqrt{n}}$. Hence, 
\begin{eqnarray*}
\inf_{\phi_0\in\Phi}P_{D_n}(\Theta(\phi_0)\subset\text{BCS}(\tau))&=&\inf_{\phi_0\in\Phi}H\left((\frac{\tilde{q}_{\tau}}{\sqrt{n}}+\frac{\phi_{10}}{1+n})\frac{1+n}{\sqrt{n}}\right)
H\left((\frac{\tilde{q}_{\tau}}{\sqrt{n}}-\frac{\phi_{20}}{1+n})\frac{1+n}{\sqrt{n}}\right)\cr
&=&H(\tilde q_{\tau})^2+o(1)\rightarrow1-\tau.
\end{eqnarray*}
  This gives (\ref{e71}).

  On the other hand, if $\phi_{20}-\phi_{10}=o(1)$, the lower BCS for $\Theta(\phi)$ is empty with a large probability. To see this, for any fixed $q$,  the lower BCS is $A=[\bar{Y}_1\frac{n}{1+n}+\frac{q}{\sqrt{n}}, \bar{Y}_2\frac{n}{1+n}-\frac{q}{\sqrt{n}}]$. Let $\Delta_n=\phi_{20}-\phi_{10}$, then
$
P(A=\emptyset)=P((\bar{Y}_2-\bar{Y}_1)\frac{n}{1+n}<\frac{2q}{\sqrt{n}})=H(\sqrt{2}q-\sqrt{\frac{n}{2}}\Delta_n)+o(1).
$
Suppose $\sqrt{n}\Delta_n=o(1)$, then $P(A=\emptyset)\rightarrow H(\sqrt{2}q)$.
This  probability is very large  for many reasonable cut-off $q$. For example, if $q=1.96$, $H(\sqrt{2}q)=0.997.$

For a numerical illustration,  set $\phi_{10}=1, \phi_{20}=1+\Delta_n$ for a sequence of  small  $\Delta_n$ that decreases to zero, and calculate the frequency that $\Theta(\phi_0)\subset$ BCS$(0.05)$ and $A=\emptyset.$  The model is nearly point identified, and point identification is achieved when $\Delta_n=0.$ Results are reported in Table \ref{t2add}.

\end{exm}

\begin{exm}[Missing data example - \textit{continued}]
Consider again the missing data example in Section \ref{smd}, where now the true $\phi_{10}$ is $\phi_{10}=1-\Delta_n$ with $\Delta_n\rightarrow0,$ that is, the probability of missing is close to zero. So the model is close to point identification.  However, suppose we still place   priors on $\phi_1$ and $\phi_2$ and $\Theta(\phi)=[\phi_1\phi_2, \phi_1\phi_2+1-\phi_1]$ as before.  Our result shows that
\begin{equation}\label{eq8.2}
P_{D_n}(\Theta(\phi_0)\subset\Theta(\hat\phi )^{\tilde{q}_{\tau}/\sqrt{n}})\rightarrow 1-\tau
\end{equation}
 when $\tilde{q}_{\tau}$ is the $1-\tau$ quantile of the posterior of
 $$
 \sqrt{n}\max\left\{\phi_1\phi_2-\phi_1-\hat{\phi}_{1}\hat{\phi}_{2}+\hat{\phi}_{1},  \hat{\phi}_{1}\hat{\phi}_{2}-\phi_1\phi_2\right\}.
 $$
It can also be shown that the coverage (\ref{eq8.2})   holds uniformly for $\phi_0$ inside a compact parameter space.
It is also easy to see that, if $\Delta_n=o(n^{-1/2})$, then for any $\tau\in(0,1)$, the lower BCS$(\tau)$ is empty with probability approaching one.   

We illustrate the above discussions using a simulated example, where $\phi_{10}=1-\Delta_n$ for a sequence of   small $\Delta_n$.   We use the uniform priors  and compute the frequency of the events that $\Theta(\phi_0)\subset\Theta(\hat\phi )^{\tilde q_{0.05}/\sqrt{n}}$ and that the lower BCS is empty. We set $\phi_{20}=0.5$ so that $\hat{\phi}_{2}$ has the maximum possible variance. Therefore, our simulation also demonstrates how sensitive  the coverage frequency is to the variance of the point identified estimator. The frequency of coverage over 500 replications are summarized in Table \ref{t2add} below.

\begin{table}[htdp]
\caption{Frequency of BCS(0.05) coverages for near point identification}
\begin{center}

\begin{tabular}{ccc|cccc}
\hline
& &  &  & $\Delta_n$ \\
&$n$ & event & 0.1 & 0.05 & 0.01 & 0 \\
\hline
 &&&&&&\\
Interval censoring&50 &  Lower BCS$=\emptyset$ & 0.966 & 0.94 & 0.972 & 0.956 \\
 && $\Theta(\phi_0)\subset\Theta(\hat\phi )^{\tilde q/\sqrt{n}}$ & 0.952 & 0.964 & 0.952 & 0.944 \\
  &&&&&&\\
&100 &   Lower BCS$=\emptyset$ & 0.964 & 0.96 & 0.962 & 0.962 \\
 &&$\Theta(\phi_0)\subset\Theta(\hat\phi )^{\tilde q/\sqrt{n}}$ & 0.962 & 0.966 & 0.958 & 0.95 \\
\hline
& &&&&&\\
Missing data &50 &  Lower BCS$=\emptyset$ & 0.998 & 1 & 1 & 1 \\
 && $\Theta(\phi_0)\subset\Theta(\hat\phi )^{\tilde q/\sqrt{n}}$& 0.95 & 0.952 & 0.942 & 0.944 \\
& &&&&&\\
&100 &  Lower BCS$=\emptyset$ & 0.99 & 1 & 1 & 1 \\
 && $\Theta(\phi_0)\subset\Theta(\hat\phi )^{\tilde q/\sqrt{n}}$& 0.952 & 0.956 & 0.958 & 0.952 \\
\hline
  \end{tabular}

  \label{t2add}
\small

\it
The frequencies (over 500 replications) that the lower BCS is empty and that the upper BCS covers the true identified set are summarized. The length of the true identified set is $\Delta_n$. The model achieves point identification when $\Delta_n=0$.

\end{center}
\end{table}

We see that the upper BCS with 95\% credible level has the coverage probability for the true identified set close to 0.95.
 Also, the lower BCS is empty almost all the times. 

 \end{exm}

\section{Financial Asset Pricing}\label{s_Financial_asset_pricing}
We develop a detailed application in financial asset pricing model, where the identified set is of direct interest.

\subsection{The model}
Asset pricing models state that the equilibrium price $P_{t}^{i}$ of a financial asset $i$ is equal to
$$
      P_{t}^{i} =  E [M_{t+1}P_{t+1}^{i}|\mathcal{I}_{t}],\qquad i=1,\ldots,N
$$

\noindent where $P_{t+1}^{i}$ denotes the price of asset $i$ at the period $(t+1)$, $M_{t+1}$ is the stochastic discount factor (SDF, hereafter) and $\mathcal{I}_{t}$ denotes the information set at time $t$. In vectorial form this rewrites as
   \begin{equation}\label{eq9.1}
      \iota =  E [M_{t+1}R_{t+1}|\mathcal{I}_{t}]
    \end{equation}

\noindent where $\iota$ is the $N$-dimensional vector of ones and $R_{t+1}$ is the $N$-dimensional vector of gross asset returns at time $(t+1)$: $R_{t+1} = (r_{1,t+1},\ldots, r_{N,t+1})^{T}$ with $r_{i,t+1} = P_{t+1}^{i}/P_{t}^{i}$. This model can be interpreted as a model of the SDF $M_{t+1}$ and may be used to detect the SDFs that are compatible with asset return data. Hansen and Jagannathan (1991) have obtained a lower bound on the volatility of SDFs that could be compatible with a given SDF-mean value and a given set of asset return data. Therefore, the set of SDFs $M_{t+1}$ that can price existing assets generally form a proper set.\\
\indent Let $m$ and $\Sigma$ denote, respectively, the vector of unconditional mean returns and covariance matrix of returns of the $N$ risky assets, that is, $m =  E (R_{t+1})$ and $\Sigma =  E (R_{t+1} - m)(R_{t+1} - m)^{T}$. Denote $\mu =  E (M_{t+1})$ and $\sigma^{2} = Var(M_{t+1})$, which are partially identified. We assume that $m$, $\Sigma$, $\mu$ and $\sigma^{2}$ do not vary with $t$.  Hansen and Jagannathan (1991) showed that given $(m,\Sigma)$, which are point identified by the observed $\{R_{t+1}\}$, if the SDF $M_{t+1}$ satisfies (\ref{eq9.1}), then its variance $\sigma^2$ should be no smaller than:
     \begin{eqnarray}\label{eq7.1}
      \sigma_{\phi}^{2}(\mu) & = & (\iota - \mu m)^{T}\Sigma^{-1} (\iota - \mu m) \equiv \phi_{1}\mu^{2} - 2\phi_{2}\mu + \phi_{3}\cr
      & & \nonumber\\
      \textrm{with } \phi_{1} & = & m^{T}\Sigma^{-1}m, \qquad \phi_{2} = m^{T}\Sigma^{-1}\iota, \qquad \phi_{3} = \iota^{T}\Sigma^{-1}\iota.
    \end{eqnarray}

\noindent Therefore, an SDF correctly prices an asset only if, for given $(m,\Sigma)$, its mean $\mu$ and variance $\sigma^{2}$ are such that $\sigma^{2}\geq \sigma_{\phi}^{2}(\mu)$, and in this case the SDF is called \textit{admissible}. Inadmissible SDFs do not satisfy  model (\ref{eq9.1}).

Define the set of admissible SDF's means and variances:
    \begin{equation*}
      \Theta(\phi) = \left\{(\mu,\sigma^{2})\in \Theta; \: \sigma_{\phi}^{2}(\mu) - \sigma^{2} \leq 0\right\}
    \end{equation*}

\noindent where $\phi = (\phi_{1},\phi_{2},\phi_{3})^{T}$ and $\Theta\subset \mathbb{R}_{+}\times\mathbb{R}_{+}$ is a compact set that we can choose based   on experiences. Usually, we can fix upper bounds $\bar{\mu}>0$ and $\bar{\sigma}>0$ as large as we want and take $\Theta = [0,\bar{\mu}] \times [0,\bar{\sigma}^{2}]$. In practice, $\bar{\mu}$ and $\bar{\sigma}$ must be chosen sufficiently large such that $\Theta(\phi)$ is non-empty. We also point out that to be consistent with our developed theory, the parameter space is chosen to be compact. Thus, the space for $\sigma^2$ includes zero. In practice, one can require $\sigma^2\geq \epsilon$ for a sufficiently small $\epsilon>0.$ For simplicity, we keep the current parameter space for $\sigma^2$, which is also used sometimes in the literature. Making inference on $\Theta(\phi)$ allows to check whether a family of SDF (and then a given utility function) prices a financial asset correctly or not. Frequentist inference for this set is carried out in Chernozhukov, Kocatulum and Menzel (2012).

 Using our previous notation,  we define $\theta = (\mu,\sigma^{2})$ and
    \begin{displaymath}
      \Psi(\theta,\phi) = \sigma_{\phi}^2(\mu) - \sigma^{2},
    \end{displaymath}
which gives a moment inequality model.

\subsection{Support function}

In this case $\Psi(\theta,\phi)$ is convex in $\theta$. More precisely, $\Psi(\theta,\phi)$ is linear in $\sigma^{2}$ and strictly convex in $\mu$ (because $\Sigma$ is positive definite so  $\phi_{1}>0$). Assumptions \ref{ass5.1}- \ref{ass5.5}  are easy to verify except for Assumptions \ref{ass5.6} and \ref{ass5.5}\textit{(i)} and \textit{(iv)}. However, it can be shown that the support function is differentiable at $\phi_{0}$ without Assumption \ref{ass5.6} being satisfied. So, our Bayesian analysis on the support function of $\Theta(\phi)$ still goes through. Assumption \ref{ass5.5} \textit{(i)} and \textit{(iv)} must be checked case by case (that is, for every region of values of $\nu$) since $\lambda(\nu,\phi)$ takes a different expression in each case, see Appendix \ref{ss_support_set_HJ} in the supplementary material.

We can rewrite the support function to be $S_{\phi}(\nu)=\Xi(\nu,\phi)^T\nu$, where
    \begin{eqnarray*}
      \Xi(\nu,\phi) & = & \arg\max_{\theta\in\Theta}\left\{\nu^T\theta; \:\Psi(\theta,\phi)\leq 0\right\}\\
      & = & \arg\max_{0\leq\mu<\bar{\mu},\:0<\sigma^{2}<\bar{\sigma}^{2}}\left\{\nu_{1}\mu + \nu_{2}\sigma^{2} - \lambda(\nu,\phi)(\phi_{1}\mu^{2} - 2\phi_{2}\mu + \phi_{3} - \sigma^{2})\right\}
    \end{eqnarray*}

\noindent where $\nu=(\nu_{1},\nu_{2})$. The support function and $\Xi(\nu,\phi)$ have explicit expressions, but they are very long and complicated. Thus, we present them in Appendix \ref{ss_support_set_HJ}.

\subsection{Dirichlet process prior}
 Let $F$ denote a probability distribution. The Bayesian model is $R_t|F\sim F$ and $  (m, \Sigma)=(m(F), \Sigma(F))$, where
    $$
m(F)=\int r F(dr),\quad \Sigma(F)=\int rr^TF(dr)-\int rF(dr)\int rF(dr)^T.
    $$
    Let us impose a Dirichlet process prior for $F$, with parameter $v_0$ and base probability measure $Q_0$ on $\mathbb{R}^N$. By Sethuraman (1994)'s decomposition, the Dirichet process prior induces a prior for $(m, \Sigma)$ as:
    $m=\sum_{j=1}^{\infty}\alpha_j\xi_j$,  and $\Sigma=\sum_{j=1}^{\infty}\alpha_j\xi_j\xi_j^T-\sum_{i=1}^{\infty}\alpha_i\xi_i\sum_{j=1}^{\infty}\alpha_j\xi_j^T$
    where $\xi_j$ are independently sampled from $Q_0$;  $\alpha_j=u_j\prod_{l=1}^j(1-u_l)$ with  $\{u_i\}_{i=1}^n$ drawn from Beta$(1,v_0)$. These priors then induce a prior for $\phi.$ The posterior distribution for $(m,\Sigma)$ can be calculated explicitly:
    \begin{eqnarray*}
 \Sigma|D_n&\sim&(1-\gamma)\sum_{j=1}^{\infty}\alpha_j\xi_j\xi_j^T+\gamma\sum_{t=1}^n\beta_tR_tR_t^n\cr
 &&-\left( (1-\gamma)\sum_{j=1}^{\infty}\alpha_j\xi_j+\gamma\sum_{t=1}^n\beta_tR_t\right)\left( (1-\gamma)\sum_{j=1}^{\infty}\alpha_j\xi_j+\gamma\sum_{t=1}^n\beta_tR_t\right)^T,
 \end{eqnarray*}
 $$
 m|D_n\sim (1-\gamma)\sum_{j=1}^{\infty}\alpha_j\xi_j+\gamma\sum_{t=1}^n\beta_tR_t,\quad \gamma\sim\text{Beta}(n,v_0),\quad \{\beta_j\}_{j=1}^n\sim Dir(1,...,1).
 $$
We can set a  truncation $K>0$, so the infinite sums in the posterior representation  are  replaced with a truncated sum.     We can then simulate the posterior for $\phi$  based on the distributions of $\Sigma|D_n$, $m|D_n$ and (\ref{eq7.1}).

\subsection{Simulation}
We present a simple simulated example. The returns $R_t$ are generated from  a 2-factor model:
$R_t=\Lambda f_t+u_t+2\iota$, where $\Lambda$ is a $N\times 2$ matrix of factor loadings. The error terms $\{u_{it}\}_{i\leq N, t\leq n}$ are i.i.d. uniform U$[-2,2]$.  Besides, the components of $\Lambda$ are standard normal, and the factors are also uniform U$[-2,2]$.  The true $m=ER_t=2\iota$, $\Sigma=\Lambda\Lambda^{T}+I_N$. 

We set $N=5, n=200$. When   the posterior is calculated,  the DGP's distributions and the factor model structure are treated unknown, and we  apply the nonparametric Dirichlet Process prior on the CDF of $R_t-m$, with parameter $v_0=3$, and based measure $Q_0=\mathcal{N}(0,1)$. We use a uniform prior for $(\sigma^2,\mu)$, and obtain the posterior distributions for $(m,\Sigma, \phi_1,\phi_2,\phi_3,\sigma^2,\mu)$.  More concretely, the prior is assumed to be:
$$
\pi(\sigma^2,\mu|\phi)=\pi(\sigma^2|\phi,\mu)\pi(\mu);\quad \pi( \sigma^2|\phi, \mu)\sim U[\sigma^2_{\phi}(\mu), \bar{\sigma}^2], \pi(\mu)\sim U[0,\bar{\mu}],
$$
where $\mu$ and $\phi$ are a priori independent. We sample 1,000 draws from the posterior of $(\phi, \sigma^2,\mu)$.  Each time we first draw $(m,\Sigma)$ from their marginal posterior distributions, based on which obtain the posterior draw of $\phi$ from (\ref{eq7.1}). In addition, draw  $\mu$ uniformly from $[0,\bar{\mu}]$, and finally $\sigma^2$ uniformly from $[\sigma^2_{\phi}(\mu), \bar{\sigma}^2]$, where $\sigma^2_{\phi}(\mu)$ is calculated based on the drawn $\phi$ and $\mu$.

\begin{figure}[htbp]
\begin{center}
\caption{ Posterior estimates of support function. Left panel is for $\nu_2\in [0,1]$, $\nu_1=\sqrt{1-\nu_2^2}$; right panel is for $\nu_2\in [-1,0]$, $\nu_1=-\sqrt{1-\nu_2^2}$}
\includegraphics[width=8cm]{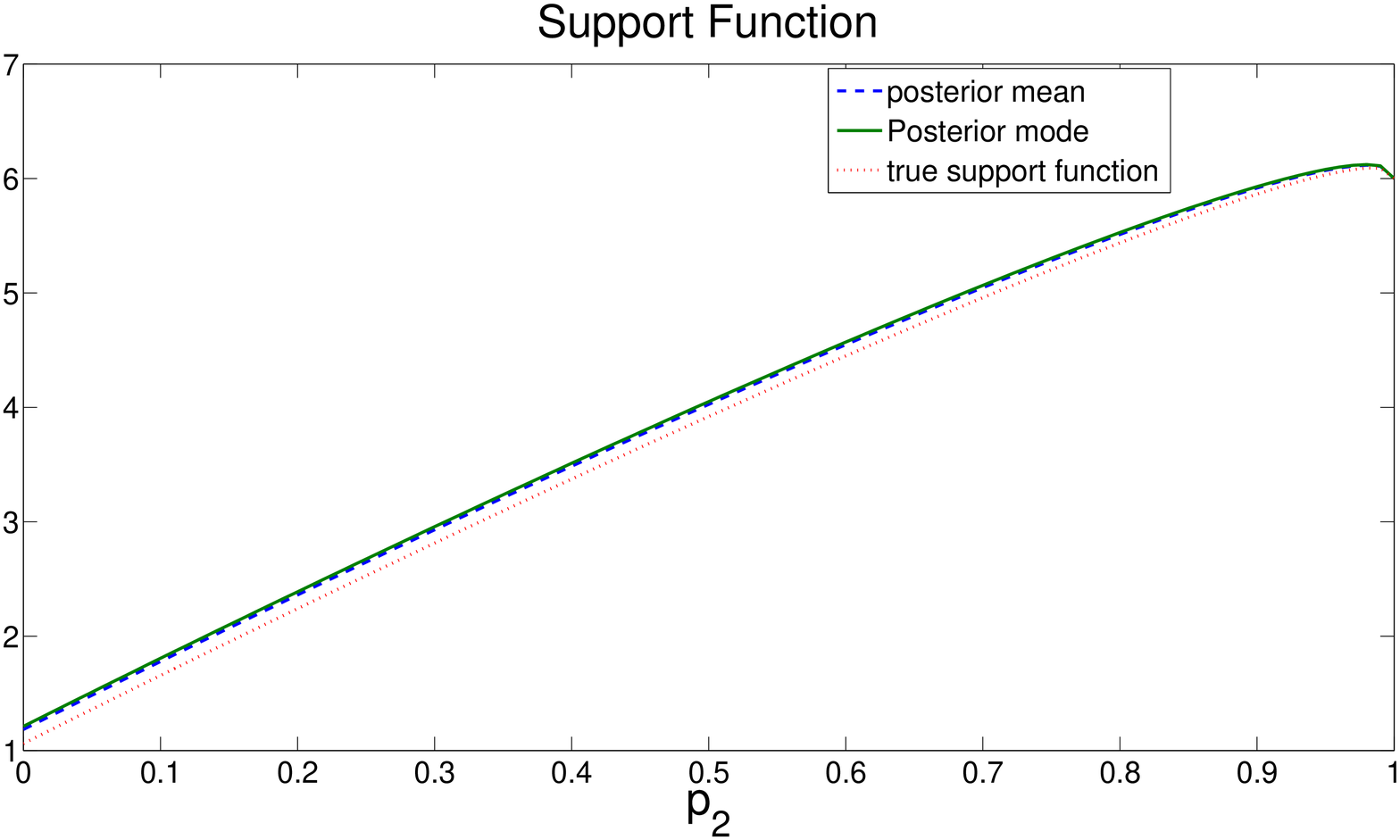}
\includegraphics[width=8cm]{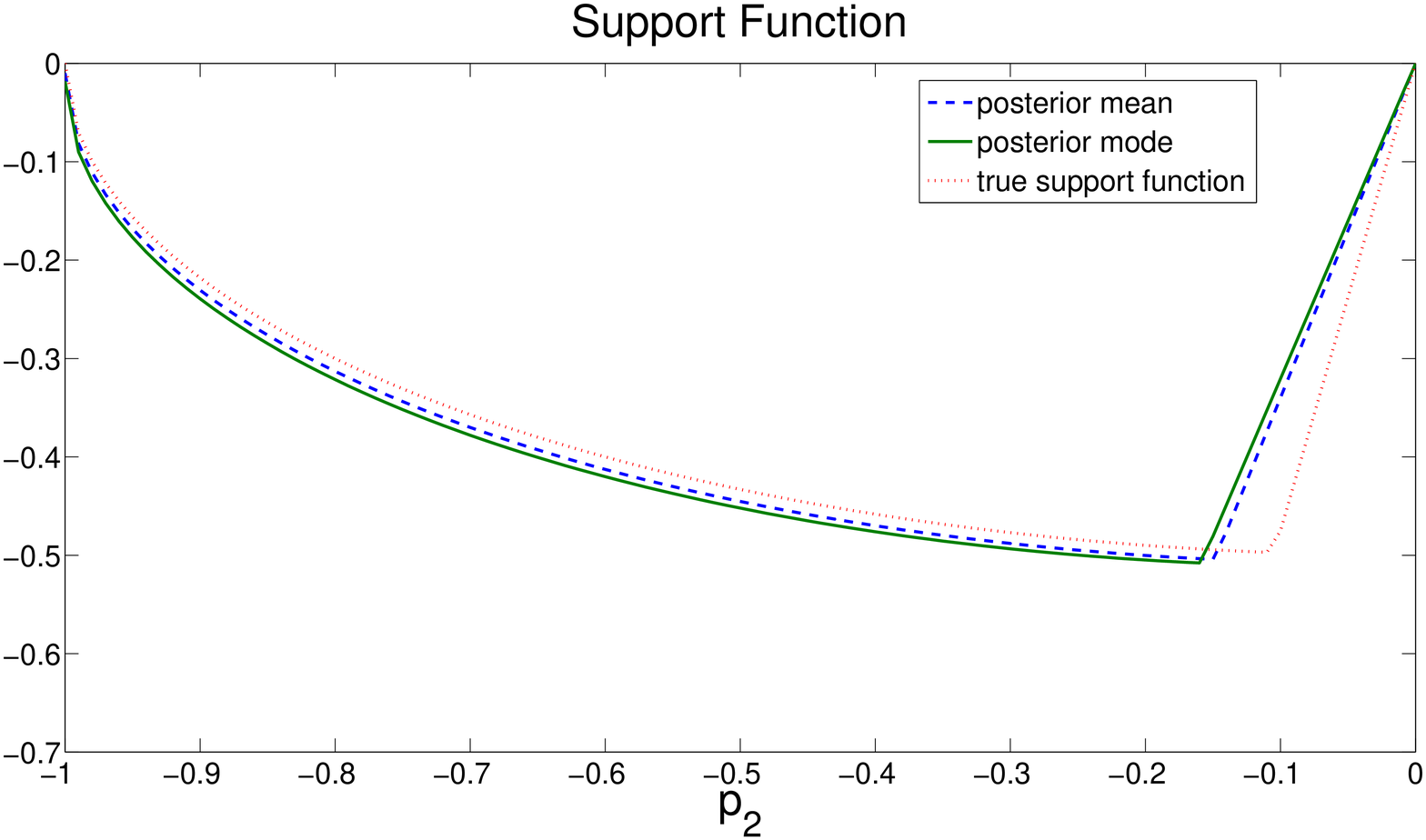}
\label{f1}
\end{center}
\end{figure}

The posterior mean $(\hat\phi_1,\hat\phi_2,\hat\phi_3)$ of $\phi$ is calculated, based on which we calculate a Bayesian estimate of the boundary of the identified set (we set $\bar{\mu}=1.4$ and $\bar{\sigma}^2=6$):
$$
\partial\Theta(\hat\phi)=\{\mu\in[0,\bar{\mu}], \sigma^2\in[0,\bar{\sigma}^2]: \sigma^2=\hat\phi_1\mu^2-2\hat\phi_2\mu+\hat\phi_3\},
$$
which is helpful to compute the BCS for the identified set. In addition, we estimate the support function $S_{\phi}(\nu)$ using  the posterior mean of $\phi$. 
  In   Figure \ref{f1}, we plot the Bayesian estimates of the support function for two cases: $\nu_2\in [0,1]$, $\nu_1=\sqrt{1-\nu_2^2}$, and $\nu_2\in [-1,0]$, $\nu_1=-\sqrt{1-\nu_2^2}$.

\begin{figure}[htbp]
\begin{center}
\caption{1,000 posterior draws of $(\mu,\sigma^2)$. Solid line is the boundary of the true identified set; dashed line represents the estimated boundary using the posterior mean; dotted line gives the 95\% BCS of the identified set. Plots are obtained based on a single set of simulated data. The BCS also covers a part of negative values for $\sigma^2$. In practice, we can truncate it to ensure it is always positive.}
\includegraphics[width=7cm]{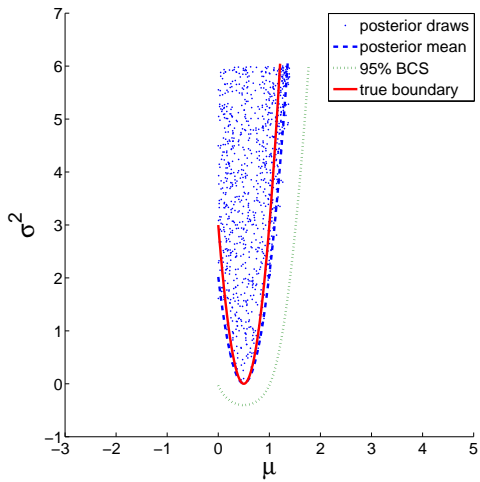}
\label{f2}
\end{center}
\end{figure}

 Using the  support function, we  calculate the 95\% posterior quantile $q_{\tau}$ for $J(\phi)$, based on which we construct the  BCS $\Theta(\hat\phi )^{q_{\tau}/\sqrt{n}}$ for the identified set. The boundary of $\Theta(\hat\phi )^{q_{\tau}/\sqrt{n}}$ is given by
 $$
\partial \Theta(\hat\phi )^{q_{\tau}/\sqrt{n}}= \left\{\mu\in[0,\bar{\mu}], \sigma^2\in[0,\bar{\sigma}^2]: \inf_{z}\sqrt{|z-\mu|^2+|\sigma^2_{\hat\phi }(z)-\sigma^2|^2}=q_{\tau}/\sqrt{n}\right\}.
 $$
In Figure \ref{f2}, we plot the posterior draws of $(\mu, \sigma^2)$, $\partial\Theta(\hat\phi), \partial \Theta(\hat\phi )^{q_{\tau}/\sqrt{n}}$ and the boundary of the true identified set. The scatter plot of posterior draws, the estimated boundaries and the BCS show that the true identified set is well estimated.

\section{Conclusion}\label{s_Conclusion}

We propose a semi-parametric Bayesian procedure for inference about partially identified models.  Bayesian approaches are appealing in many aspects.  Classical Bayesian approach in this literature has been assuming a known  likelihood function. However, in many applications econometric models   only identify a set of moment inequalities, and therefore assuming a known likelihood function suffers from the risk of misspecification, and may result in inconsistent estimations of the identified set. On the other hand, Bayesian approaches that use moment-condition-based likelihoods such as the limited information and exponential tilted empirical likelihood, though guarantee the consistency, lack of probabilistic interpretations, and do not provide an easy way to make inference about the identified set.  Our approach, on the contrary, only requires a set of moment conditions but still possesses a pure Bayesian interpretation. Importantly, we can conveniently analyze both the partially identified parameter and its identified set. Moreover, we shed light on many appealing features of our proposed approach such as the computational efficiency for subset inference and projections.

Our analysis primarily focuses on identified sets which are closed and convex. These sets are completely characterized by their support function. By imposing a prior on the support function, we construct its posterior distribution. It is shown that the support function for a very general moment inequality model admits a linear expansion, and its posterior is consistent. The Bernstein-von Mises theorem for the posterior of the support function is proven.

Some researcher may argue that frequentist and Bayesian approaches are asymptotically equivalent because the prior is ``washed away" as the sample size increases. So why bother with Bayesian asymptotics? Interestingly,   this is no longer the case under partial identification.  As was shown by Moon and Schorfheide (2012), when making inference about the partially identified parameter, the BCS can be strictly smaller than the FCS.   Therefore, the two fundamental statistical methodologies provide different inferences that are not necessarily equivalent even asymptotically.  This paper completes such a comparison. We also establish the large-sample correspondence of the two approaches for the identified set. It is also illustrated that the results hold in the uniform sense, in particular when point identification is nearly achieved.


\newpage

\appendix
\section*{Appendix}
\section{Semi-parametric prior}\label{sss_semiparametric_prior}


 We present in this section a prior scheme for $(\phi,F)$ which is an alternative to the prior scheme presented in section \ref{sss_nonparametric_prior}. The idea is to reformulate  the sampling distribution $F$ in terms of $\phi$ and a nuisance parameter $\eta\in\mathcal{P}$, where $\mathcal{P}$ is an infinite-dimensional measurable space. Therefore, $F=F_{\phi, \eta}\in\mathcal{F}=\{F_{\phi,\eta};\phi\in\Phi,\: \eta\in \mathcal{P}\}$.  
We denote by $l_{n}(\phi,\eta)$ the model's likelihood function. One of the appealing features of this semi-parametric prior is that it allows us to impose a prior $\pi(\phi)$ directly on the identified parameter $\phi$, which is convenient whenever we have good prior information regarding $\phi$. 

For example, in the interval censored data example (Example \ref{ex2.1}), we can write
$$
Y_1=\phi_1+u,\quad Y_2=\phi_2+v
$$
where $(u, v)$ are independent random variables with zero means and have unknown densities $(\eta_1,\eta_2)$.
Then $\eta=(\eta_1, \eta_2)$, and the likelihood function is
$$
l_n(\phi,\eta)=\prod_{i=1}^n\eta_1(Y_{1i}-\phi_1)\eta_2(Y_{2i}-\phi_2).
$$
We put priors on $(\phi, \eta)$.   Examples of priors on density functions $\eta_1$ and $\eta_2$ include mixture of Dirichlet process priors, Gaussian process priors, etc. (see Ghosal et al. (1999) and Amewou-Atisso et al. (2003)). We list some commonly used priors for densities in the examples below.

We place an independent prior as $\pi(\phi, \eta)=\pi(\phi)\pi(\eta)$. Then the joint prior distribution $\pi(\theta,\phi, \eta)$ is naturally decomposed as
  $
      \pi(\theta,\phi, \eta) = \pi(\theta|\phi)\pi(\phi)\pi(\eta)
$
and the Bayesian experiment is
    \begin{displaymath}
      X |\phi,\eta \sim F_{\phi,\eta},\qquad (\phi,\eta)\sim  \pi(\phi)\pi(\eta), \qquad \theta|\phi, \eta\sim \pi(\theta|\phi).
    \end{displaymath}

The posterior distribution of $\phi$ has a density function given by
\begin{equation}\label{eq2.5}
p(\phi|D_n)\propto\int_{\mathcal{P}}\pi(\phi,\eta)l_n(\phi,\eta)d\eta.
\end{equation}
Then the marginal posterior of $\theta$ is, for any measurable set $B\subset\Theta$:
\begin{equation*}\label{eq_semiparametric_posterior_theta}
  P(\theta\in B |D_n)\propto \int_{\Phi}\pi(\theta\in B|\phi)p(\phi|D_n)d\phi=E[\pi(\theta\in B|\phi)|D_n]
\end{equation*}
where the conditional expectation is taken with respect to the posterior of $\phi$.  Moreover,  the posteriors of $\Theta(\phi)$ and $S_{\phi}(\cdot)$ are deduced from that of $\phi$. Suppose for example we are interested in whether $\Theta(\phi)\cap A$ is an empty set for some $A\subset \Theta$,  then the posterior probability $P(\Theta(\phi)\cap A=\emptyset|D_n)$ is relevant.

\begin{exm}[Interval regression model - \textit{continued}] \label{ex4.1}Consider Example \ref{ex2.2}, where $\phi=(\phi_1,\phi_2,\phi_3)=(E(ZY_1), E(ZX^T), E(ZY_2)).$ Write
$
ZY_1=\phi_1+u_1, 
$
$
ZY_2=\phi_3+u_3, 
$ and
$
\text{vec}(ZX^T)=\text{vec}(\phi_2)+u_2,
$
where $u_1, u_2$ and $u_3$ are correlated and their joint unknown probability density function is $\eta(u_1, u_2, u_3)$. The likelihood function is then
$$
l_n(\phi, \eta)=\prod_{i=1}^n \eta(Z_iY_{1i}-\phi_1,Z_iY_{2i}-\phi_3,\text{vec}(Z_iX_{i}^T)-\text{vec}(\phi_2)).
$$
Many nonparametric priors can be used for $\pi(\eta)$ in a ``location-model" of this type, where the likelihood takes the form $l_n(\phi,\eta)=\prod_{i=1}^n\eta(X_i-\phi)$, including  Dirichlet mixture of normals (Ghosal et al. 1999), random Bernstein polynomials (Walker et al. 2007), and finite mixture of normals (Lindsay and Basak 1993).

\end{exm}

We provide some examples of prior for the density $\eta$.

\begin{exm} The finite mixture of normals (e.g., Lindsay and Basak (1993),  Ray and Lindsay (2005)) assumes $\eta$ to take the form
$$
\eta(x)=\sum_{i=1}^kw_ih(x; \mu_i, \Sigma_i),
$$
where $h(x; \mu_i, \Sigma_i)$ is the density of a multivariate normal distribution with mean $\mu_i$ and variance $\Sigma_i$ and $\{w_i\}_{i=1}^k$ are unknown weights such that $\sum_{i=1}^kw_i\mu_i=0.$ Then $\int\eta(x)xdx=\sum_{i=1}^kw_i\int h(x; \mu_i, \Sigma_i)xdx=0.$  We impose prior $\pi(\eta)=\pi(\{\mu_l,\Sigma_l, w_l\}_{l=1}^{k})$, then
\begin{eqnarray*}
p(\phi|D_n)\propto\int\pi(\phi)\prod_{i=1}^n\sum_{j=1}^kw_jh(X_i-\phi; \mu_j,\Sigma_j)\pi(\{\mu_l,\Sigma_l,w_l\}_{l=1}^{k})dw_jd\mu_jd\Sigma_j.
\end{eqnarray*}

\end{exm}

\begin{exm}\label{ex2.6}
Dirichlet mixture of normals (e.g., Ghosal \textit{et al.} (1999) Ghosal and van der Vaart (2001), Amewou-Atisso, et al. (2003)) assumes
$$
\eta(x)=\int h(x-z;0,\Sigma)dH(z)
$$
where $h(x;0,\Sigma)$ is the density of a multivariate normal distribution with mean zero and variance $\Sigma$ and $H$ is a probability distribution such that $\int zH(z)dz=0.$ Then $\int x\eta(x)dx=0.$ To place a prior on $\eta$, we let $H$ have the Dirichlet process prior distribution $D_{\alpha}\equiv \mathcal{D}(\nu_{0},Q_{0})$ where $\alpha$ is a finite positive measure, $\nu_{0}=\alpha(\mathcal{X})\in\mathbb{R}_{+}$ and $Q_{0}=\alpha/\alpha(\mathcal{X})$ is a base probability on $(\mathcal{X},\mathcal{B}_{x})$ such that $Q_{0}(x)=0$, $\forall x\in(\mathcal{X},\mathcal{B}_{x})$. In addition, we place a prior on $\Sigma$ independent of the prior on $H$. Then
$$
p(\phi|D_n)\propto\int\pi(\phi)\pi(\Sigma)D_{\alpha}(H)\prod_{i=1}^n\int h(X_i-\phi-z;0,\Sigma)dH(z)d\Sigma dH.
$$

\end{exm}

\begin{exm}
Random Bernstein polynomials (e.g., Walker et al. (2007) and Ghosal (2001)) allow to write the density function $\eta$ as
$$
\eta(x)=\sum_{j=1}^k[H(j/k)-H((j-1)/k)]\mathcal{B}e(x;j,k-j+1),
$$
where $\mathcal{B}e(x; a,b)$ stands for the beta density with parameters $a,b>0$ and $H$ is a random distribution function with prior distribution assumed to be a Dirichlet process. Moreover, the parameter $k$ is also random with a prior distribution independent of the prior on $H$. Then
$
p(\phi|D_n)\propto\int\pi(\phi)\prod_{i=1}^n\eta(X_i-\phi)\pi(H)\pi(k)dHdk.
$
$\square$
\end{exm}

Other commonly used priors on density functions are wavelet expansions (Rivoirard and Rousseau (2012)), Polya tree priors (Lavine (1992)), Gaussian process priors (van der Vaart and van Zanten (2008), Castillo (2008)), etc.


\section{Posterior Concentration for $\phi$}\label{s_appendix_posterior_concentration_phi}


This section focuses on the case where the prior for $\phi$ is specified through the semi-parametric prior described in Appendix \ref{sss_semiparametric_prior}

Much of the literature on posterior concentration rate for Bayesian non-parametrics relies on the notion of \textit{entropy cover number}, which we now define as follows.  Recall that for i.i.d. data, the likelihood function can be written as $
l_n(\phi, \eta)=\prod_{i=1}^nl(X_i; \phi, \eta),
$
where $l(x; \phi, \eta)$ denotes  the density of the sampling distribution. Let
$$
G=\{l(\cdot; \phi, \eta): \phi\in\Phi, \eta\in\mathcal{P}\}
$$
be the family of likelihood functions. We assume $\mathcal{P}$ is a metric space with a norm $\|.\|_{\eta}$, which then induces a norm $\|.\|_G$ on $G$ such that  $\forall l(\cdot; \phi, \eta)\in G$,
$$
\|l(\cdot; \phi, \eta)\|_G=\|\phi\|+\|\eta\|_{\eta}.
$$
For instance, in the examples of intervel censoring data and interval regression, $l(x; \phi, \eta)=\eta(x-\phi)$ and  $\|\eta\|_{\eta}=\|\eta\|_1=\int |\eta(x)|dx$. Then  in this case $
\|l(., \phi, \eta)\|_G=\|\phi\|+\|\eta\|_1.
$

Define the entropy cover number $
\mathcal{N}(\rho, G, \|.\|_G)
$
to be the minimum number of balls with radius $\rho$ needed to cover $G$. The importance of the entropy cover number on nonparametric Bayesian asymptotics has been realized for a long time.  We refer the audience to Kolmogorov and Tikhomirov (1961) and van der Vaart and Wellner (1996) for good early  references.

We first present the assumptions that are sufficient to derive the posterior concentration rate for the point identified $\phi$. The first one is placed on the entropy cover number.
\begin{assum}\label{assa.1}
Suppose  for all  $n$ large enough,
$$
\mathcal{N}(n^{-1/2}(\log n)^{1/2}, G, \|.\|_G)\leq n.
$$
\end{assum}
This condition requires that the ``model" $G$ be not too big. Once this condition holds, then for all $r_n\geq n^{-1/2}(\log n)^{1/2}$, $\mathcal{N}(r_n, G, \|.\|_G)\leq \exp(nr_n^2).$ Moreover, it ensures the existence of certain tests as given in Lemma \ref{la.2} in the supplementary appendix, and hence it can be replaced by the test condition that are commonly used in the literature of posterior concentration rate, e.g., Jiang (2007),  Ghosh and Ramamoorthi  (2003). The same condition has been imposed by Ghosal et al. (2000) when considering Hellinger rates, and Bickel and  Kleijn (2012) when considering semi-parametric posterior asymptotic normality, among others. When the true density $\eta_0$ belongs to the family of location mixtures, this condition was verified by Ghosal et al. (1999, Theorem 3.1).

The next assumption places conditions on the prior for $(\phi,\eta)$. For each $(\phi, \eta)$, define
$$
K_{\phi, \eta}=E\left[\log\frac{l(X; \phi_0,\eta_0)}{l(X; \phi, \eta)}\bigg|\phi_0,\eta_0\right]=\int \log\left(\frac{l(x; \phi_0,\eta_0)}{l(x; \phi, \eta)}\right)l(x;\phi_0,\eta_0)dx
$$
$$
V_{\phi,\eta}=\var\left[\log\frac{l(X; \phi_0,\eta_0)}{l(X; \phi, \eta)}\bigg|\phi_0,\eta_0\right]
=\int \log^2\left(\frac{l(x; \phi_0,\eta_0)}{l(x; \phi, \eta)}\right)l(x;\phi_0,\eta_0)dx-K_{\phi,\eta}^2.
$$
\begin{assum}\label{assa.2} The prior $\pi(\phi,\eta)$ satisfies:
$$
\pi\left(K_{\phi,\eta}\leq \frac{\log n}{n},\quad V_{\phi,\eta}\leq \frac{\log n}{n}\right)n^M\rightarrow\infty
$$
for some $M>2$.
\end{assum}
Intuitively, when $(\phi, \eta)$ is close to $(\phi_0,\eta_0)$, both $K_{\phi,\eta}$ and $V_{\phi,\eta}$ are close to zero. Hence this assumption requires that the prior have sufficient amount of support around the true point identified parameters in terms of the Kullback-Leibler distance. Such a prior condition has also been commonly imposed in the literature on semi-parametric posterior concentration, e.g., Ghosal et al. (1999 (2.10), 2000 condition 2.4), Shen and Wasserman (2001, Theorem 2) and   Bickel and  Kleijn (2012, (3.13)). Moreover, it has been verified in the literature using  the sieve  prior (Shen and Wasserman  2001), Dirichlet mixture prior (Ghosal et al. 1999) and Normal mixture prior (Ghosal and van der Vaart 2007).

We are now ready to present the posterior concentration rate for $\phi$, whose proof is given in  Appendix \ref{s_appendix_lemmas_posterior_concentration_phi} of the supplementary appendix.

\begin{thm}\label{ta} Suppose the data $X_1,...,X_n$ are i.i.d. Under Assumptions \ref{assa.1} and \ref{assa.2}, for some $C>0$,
$$
P(\|\phi-\phi_0\|\leq Cn^{-1/2}(\log n)^{1/2}|D_n)\rightarrow^p1.
$$
\end{thm}

\vspace{3em}

All the proofs are given in the supplementary appendix.


\end{document}